\documentclass[nonacm=true,format=acmsmall]{acmart}

\acmJournal{PACMPL}
\acmVolume{1}
\acmNumber{ICFP} 
\acmArticle{1}
\acmYear{2023}
\acmMonth{1}
\acmDOI{} 
\startPage{1}

\setcopyright{none}

\usepackage{booktabs}   
\usepackage{subcaption} 

\usepackage{amsmath, listings}
\usepackage{csvsimple}
\usepackage{xstring}
\usepackage{graphicx}
\graphicspath{ {./images/} }
\usepackage[ruled,vlined,linesnumbered,commentsnumbered]{algorithm2e}
\usepackage{xcolor}
\usepackage{float}
\usepackage{verbatim}
\usepackage{longtable}
\usepackage{multirow, multicol}
\usepackage{subfiles}

\definecolor{codegreen}{rgb}{0,0.6,0}
\definecolor{codegray}{rgb}{0.5,0.5,0.5}
\definecolor{codepurple}{rgb}{0.58,0,0.82}
\definecolor{backcolour}{rgb}{0.95,0.95,0.92}

\lstdefinestyle{mystyle}{
    backgroundcolor=\color{backcolour},   
    basicstyle=\ttfamily\footnotesize,
    breakatwhitespace=false,         
    breaklines=true,                 
    captionpos=b,                    
    keepspaces=true,                 
    numbers=left,                    
    numbersep=5pt,                  
    showspaces=false,                
    showstringspaces=false,
    showtabs=false,                  
    tabsize=2
}
\newcommand{\alt}{~~|~~}

\newcommand{\krakenSpace} {\textit{Kraken} }
\newcommand{\kraken} {\textit{Kraken}}

\newcommand{\falsev}          {\mathsf{false}}
\newcommand{\truev}           {\mathsf{true}}

\newcommand{\Ctxt}            {\mathcal{E}}
\newcommand{\InCtxt}  [1]     {\Ctxt[#1]}

\newcommand{\kenv}     [3] {\langle \langle #1~|#2,~#3 \rangle \rangle}

\newcommand{\kprim}    [2] {\langle #1~\textbf{#2} \rangle}
\newcommand{\kcomb}    [5] {\langle \textbf{comb} ~ #1 ~ #2 ~ #3 ~ #4 ~ #5\rangle}
\newcommand{\keval}    [2] {[\text{eval} ~ #1 ~ #2]}
\newcommand{\kcombine} [3] {[\text{combine} ~ #1 ~ #2 ~ #3 ]}

\newcommand{\kpeval}      [5] {[\text{peval} ~ #1 ~ #2 ~ #3 ~ #4 ~ #5]}

\newcommand{\kmark}   [1] {\text{mark}(#1)}
\newcommand{\kmkd}    [2] {/#1/#2}
\newcommand{\kval}        {\mathsf{val}}
\newcommand{\kfresh}      {\mathsf{freshCall}}
\newcommand{\katt}    [2] {\mathsf{atmdCall}~#1~#2}
\newcommand{\kunval}  [1] {\text{unval}(#1)}
\newcommand{\kprog}   [1] {\text{neededIDs}(#1)}
\newcommand{\kuprog}  [1] {\text{upperIDs}(#1)}
\newcommand{\kiprog}  [1] {\text{resumeForms}(#1)}
\newcommand{\kpval}     [4] {[\text{peval} ~ #1 ~ #2 ~ #3 ~ #4]}
\newcommand{\kpcombine} [5] {[\text{combine} ~ #1 ~ #2 ~ #3 ~ #4 ~ #5]}
\newcommand{\kunder}    [5] {[\text{under} ~ #1 ~ #2 ~ #3 ~ #4 ~ #5]}

\begin{document}
\title[Kraken]{Practical compilation of fexprs using partial evaluation}
\subtitle{Fexprs can performantly replace macros in purely-functional Lisp}


\author{Nathan Braswell}
\affiliation{
  \position{PhD Student}
  \department{School of Computer Science}       
  \institution{Georgia Institute of Technology} 
  \city{Atlanta}
  \state{GA}
  \country{United States}                    
}
\email{nathan.braswell@gtri.gatech.edu}      

\author{Sharjeel Khan}
\affiliation{
  \department{School of Computer Science}       
  \institution{Georgia Institute of Technology} 
  \city{Atlanta}
  \state{GA}
  \country{United States}                    
}
\email{smkhan@gatech.edu}      

\author{Santosh Pande}
\affiliation{
  \department{School of Computer Science}       
  \institution{Georgia Institute of Technology} 
  \city{Atlanta}
  \state{GA}
  \country{United States}                    
}
\email{santosh.pande@cc.gatech.edu}      

\begin{abstract}

Macros are a common part of Lisp languages, and one of their most lauded features.
Much research has gone into making macros both safer and more powerful resulting in developments in multiple areas, including maintaining hygiene, and typed program staging \cite{rompf2013optimizing}.
However, macros do suffer from various downsides, including being second-class. Particularly egregious for eager functional programming, they are unable to be passed to higher-order functions or freely composed.

  Fexprs, as reformulated by John Shutt \cite{shutt2010fexprs}, provide a first-class and more powerful alternative to macros that meshes well with pure functional programming.
 Unfortunately, naive execution of fexprs is much slower than macros due to re-executing unoptimized operative combiner code at runtime that, in a macro-based language, would have been expanded and then optimized at compile time.
To show that fexprs can be practical replacements for macros, we formulate a small purely functional fexpr based Lisp, \kraken, with an online partial evaluation and compilation framework that supports first-class, partially-static-data environments and can completely optimize away fexprs that are used and written in the style of macros.
We show our partial evaluation and compilation framework produces code that is more than 70,000 times faster than naive interpretation due to the elimination of repeated work and exposure of static information enabling additional optimization.
  In addition, our \krakenSpace compiler performs better compared to existing interpreted languages that support fexprs, including improving on NewLisp's~\cite{mueller2018newlisp} fexpr performance by 233x on one benchmark.

\end{abstract}



\maketitle

\section{Introduction}
\label{sec:intro}

Lisps languages~\cite{winston1986lisp,plt-tr1,hickey2008clojure} generally have two different abstraction methods: functions and macros. These two abstractions differ in their characteristics and semantics.
Functions operate at run-time and always evaluate their parameters, while macros operate solely at expansion time and do not evaluate their parameters.
Functions can sometimes (depending on the implementation) be used within macros but with restrictions.
These restrictions split the language into two, and do not exhibit a few of the key tenets of functional programming - namely, being higher-order and supporting composition.

Macro systems generally attempt to be hygienic by either preventing or making it difficult to manipulate the code environment of the macro expansion \cite{clinger1991macros}.
However, manipulation of the code environment is often needed in special cases.
The solution is typically various escape hatches that complicate the macro system.
For these reasons, practical macro systems are often fairly complex and different from the language they are embedded within.
In addition, because macros are expanded away, debuggability suffers as programmers must mentally expand macros to determine the cause of an error in code they did not explicitly write, but was generated by macro expansion. 

A possible solution to these problems, fexprs, were created as a first-class and more powerful alternative in the 1960s and reformulated in 2010~\cite{shutt2010fexprs}.
Following Shutt's terminology, fexprs unify functions, macros, and built-in language forms into a single concept called a combiner.
Where \textit{lambda} introduces functions, \textit{vau} introduces combiners.
A combiner is able to evaluate its arguments 0 or more times in the calling environment, and additionally receives that environment as a parameter, allowing it to dynamically evaluate code in the same scope from which it was called.
Combiners that evaluate their parameters 0 times are called operatives, while those that evaluate their parameters 1 or more times are called applicatives.
Applicatives that evaluate their parameters one time and do not use the calling environment are simply normal functions.
Operatives can subsume macros, built in special forms, and even control flow, like \textit{if} and \textit{lambda} itself.

A programming language based on combiners, using operatives instead of macros, provides a range of benefits:
\begin{itemize}
    \item \textbf{Simpler Language}: While not having a separate macro system already makes for a simpler language, supporting operative combiners as first-class values further simplifies the definition of the language because there are no special forms.
      For instance, \textit{if} and \textit{vau} are just built-in operative combiners with support from the language definition or hypothetical interpreter loop.
      Other language features often implemented via macros like \textit{let}, \textit{and}, \textit{or}, \textit{cond} can of course be implemented as derived operatives, and going further, even \textit{lambda} can be derived instead of built-in.
    
    \item \textbf{Simpler Mental Model and Better Debugging}: By unifying functions and macros (and formerly special forms) into one concept, debugging and mentally following macro-like operatives can be the same as for normal functions. When encountering an error, one can look at a stack trace and use a debugger to inspect stack variables to figure out what went wrong. The prototype debugger we developed in \krakenSpace can show this information even when it would otherwise be optimized away by re-evaluating the side-effect-free code necessary to reconstruct the missing information. In a language based on macros one would need to print out different expansions of the macro and then try to figure out which one failed and why, without debugger support.
    
    \item \textbf{Greater Flexibility and Expressivity}: All combiners, including operatives, are first class and can thus be named, passed to higher-order combiners, composed, or put into data-structures.
      While many languages have first class functions, first class macros have not enjoyed the same success.
      Combiners are both in one.
      For example, in Scheme \textit{and} is often a macro that expands to conditional control flow, meaning that it cannot be passed to a higher-order function such as \textit{fold} without first being wrapped in a \textit{lambda}, as seen in Listing~\ref{code:and1}.
      When \textit{and} is an operative, it can be freely passed to higher-order combiners as is, as demonstrated in Listing~\ref{code:and2}.

    \begin{lstlisting}[language=Lisp,caption={Scheme's version of \textit{and} example},label=code:and1]
        > (fold and #t (list #t #t))
        Exception: invalid syntax and 
        > (fold (lambda (a b) (and a b)) #t (list #t #t))
        #t\end{lstlisting}
    \begin{lstlisting}[language=Lisp,caption={\kraken's version of \textit{and} example},label=code:and2]
        > (foldl and true (array true true))
        true\end{lstlisting}

    The power of fexprs is reminiscent of advanced macro systems like that of Racket~\cite{plt-tr1}, which advocates for the definition of entirely new languages using its impressive, but complex macro system.
    A language using fexprs can have similar expressive power, but with simpler semantics.
    Listing~\ref{code:express} shows a sample fexpr implementation of \textit{let1} (a simple version of a let binding supporting only one variable) and \textit{lambda} (using \textit{let1}), demonstrating the early stages of bootstrapping a full language out of an extremely spartan base language where \textit{vau} is the only abstraction operator.
    The details of how this works will be explained later in the paper.
    This is only to give the flavor of how features most languages would consider primitive and built-in can instead be defined inside the language itself:

      \begin{lstlisting}[language=Lisp,caption={Fexpr implementation of \textit{let1} and \textit{lambda}},label=code:express]
        ((wrap (vau (let1)
        ; Definition of lambda
        (let1 lambda (vau se (p b1)
              (wrap (eval (array vau p b1) se)))
          ; a simple function that multiplies its argument by two
          (lambda (n) (* n 2))
        )
        ; Definition of let1
        )) (vau de (s v b)
              (eval (array (array vau (array s) b) (eval v de))
                    de)))\end{lstlisting} 
\end{itemize}

Despite all these benefits, naive execution of a pure language based on fexprs is exceedingly slow.
{\it During its evaluation, the body of the of the called combiner is re-executed every time it is invoked, not just for function-like calls to applicatives, but for all macro-like calls to operatives too}.  
This re-execution happens for ever combiner call in the definition of the called operative as well.
As a result, the re-executions will cause slowdowns likely to be exponential in the depth of the chain of definitions of macro-like operatives using other macro-like operatives in their definition.
On the other hand, a macro in a macro system would have been executed once at expansion time and never during runtime.
In the case of macros used in the definition of other macros, they will all be completely expanded before compilation. 
Because it is impossible to tell from syntax alone whether an argument to a combiner is evaluated or passed unevaluated (because it is impossible to tell if the call is going to be to an applicative or operative), code with fexprs is difficult to compile and optimize.
As a result, typical implementations of fexprs leave them unoptimized and entirely interpreted, augmenting the slowness issue.

Some languages like NewLisp and PicoLisp \cite{mueller2018newlisp, burger2013picolisp} that implemented some part of fexprs generally limit the number of layered fexprs to avoid compounding slowdowns. They chose to implement many combiners directly in the interpreter for speed instead of writing them as derived fexprs. Any code using fexprs in these languages still incur performance penalties or crash after hitting a limit. Other works \cite{shutt2010fexprs, kearsleyimplementing} have described fexpr languages and shown their usefulness but the few implementations based on them have been extremely slow.
As a result, a practical (fast) language based primarily on fexprs does not exist.

One solution to the performance issues of fexprs is partial evaluation.
Partial evaluators have been developed for numerous languages~\cite{elphick2003partial, LLOYD1991217, andersen1992self,meyer1991techniques,alpuente1998partial,10.1145/582153.582181} and for different domains~\cite{futamura1971partial,berlin1990partial,berlin1990compiling}.
The purpose of partial evaluation is {\it to specialize code based on values known at partial-evaluation time in order to do less work at execution time}, hopefully improving performance~\cite{10.1145/3140587.3062381,10.1145/243439.243447}. Partial evaluation can be broken into online and offline techniques with \cite{10.1145/243439.243447} explaining the differences the best. 
In addition, John Shutt~\cite{shutt2010fexprs} also suggested partial evaluation might be a solution to fexpr performance but did not provide any specific details or implementation.
Despite well-known online and offline partial evaluation techniques~\cite{10.1145/243439.243447,10.1145/158511.158707}, there is still no partial evaluation solution for fexprs, and so fexprs remain slow in all existing implementations.


In this work, we propose the first practical (fast) purely functional Lisp language based entirely on fexprs, \kraken.
\krakenSpace utilizes an online partial evaluator specifically created to evaluate away all calls to macro-esque operatives paired with a compiler backend that takes advantage of the static information exposed by the partial evaluator to produce reasonably performant WebAssembly binaries.
Using it, we show that {\it a functional Lisp based on fexprs can be approximately as efficient and at least as expressive as one based on macros}.

\textbf{\krakenSpace Language} (\S\ref{sec:base}): Our first contribution is our purely  functional fexpr-based Lisp called \kraken. The language is based on John Shutt's definition of pure Vau calculus~\cite{shutt2010fexprs} (using fexprs and based on \textit{vau} instead of \textit{lambda}) but augmented with explicit primitive data and operations to demonstrate a more practical language.

\textbf{Partial Evaluation and Compiler Optimizations} (\S\ref{sec:partial} - \S\ref{sec:opt}): 
 Our second contribution is the tailored partial evaluation algorithm and compiler optimizations that enable an fexpr-based functional language to have competitive performance.
 Partial evaluation will evaluate away any user-defined operatives that behave like macros, and the compiler will inline any primitive operatives (if, etc), leaving only the non-macro applicatives for runtime.
 Compiler optimizations (type-inference-informed primitive inlining, single-use-closure-inlining, and lazy-environment-creation) remove many of the remaining inefficiencies in our language.
 This combined partial-evaluation and compilation technique shows that macro-esque operatives can perform as well as macros due to being compiled away statically, similar to how macros would be expanded away.

\textbf{Evaluation of the Language} (\S\ref{sec:eval}): Our final contribution is the evaluation of the language and compiler on a few benchmarks to demonstrate its practicality.
Firstly, we show partial evaluation with compiler optimizations improves the runtime performance by over \textbf{70,000x} compared to our baseline interpreter.
Secondly, we show \kraken's optimized and compiled code performs significantly better than NewLisp's interpreted fexpr~\cite{mueller2018newlisp} implementation, by \textbf{233x} in one benchmark.
Lastly, we compare our runtime performance against NewLisp's macro implementation and end up faster than it as well.

The paper begins by discussing our general compilation flow in Section 2 before defining a simplified version of our \krakenSpace language in Section 3.
Section 4 contains our core contribution, the partial evaluation algorithm focused on evaluating away macro-esque operatives.
This is followed by a discussion of the major optimizations that work hand-in-hand with the partial evaluation algorithm in Section 5.
Section 6 presents our benchmarks demonstrating the dramatic speedups our algorithm achieves over the naive interpretation of fexprs before Section 7 lists related work and Section 8 concludes.

\section{General Compilation Framework}
\label{sec:overview}

\begin{figure}[!ht]
\includegraphics[width=\textwidth]{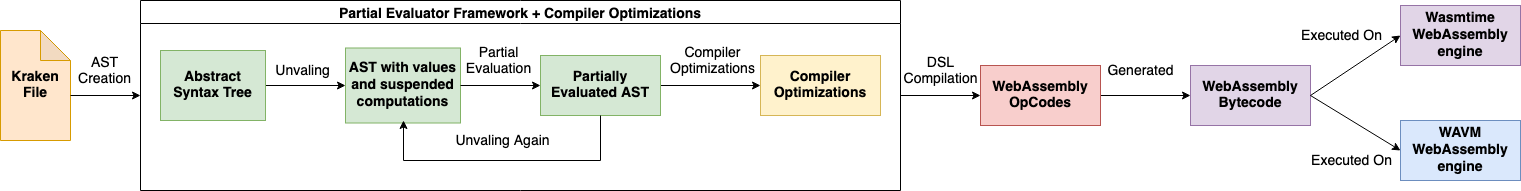}
	\caption{The Compilation Framework (including partial evaluation and compiler optimizations) compiles \krakenSpace code into a WebAssembly binary that (for our benchmarks) executes on either the WAVM WebAssembly engine or the Wasmtime WebAssembly engine}
\label{fig:overview}
\end{figure}

The \krakenSpace language gets compiled into WebAssembly before it is executed on either the Wasmtime WebAssembly engine or WAVM WebAssembly engine as seen in Fig.~\ref{fig:overview}.
Initially, the \krakenSpace code is parsed to create an abstract syntax tree (AST).
The AST gets 'marked', adding bookkeeping information needed for partial evaluation to the AST, and is then 'unvaled', annotating the sections of the AST which will be evaluated as suspended computations.
Once the AST is unvaled, the AST is partially evaluated by checking each suspended computation to see if it can be evaluated based on the current (generally partially-static-data) environment.
While the compiler backend walks the AST and emits WebAssembly bytecode, it performs optimizations like type-inference-based primitive inlining, single-use-closure inlining, lazy-environment-creation, and basic tail call elimination to further improve code quality for better performance.
This bytecode can then be executed on any compliant WebAssembly engine.

\section{\krakenSpace Language}
\label{sec:base}

As mentioned before, the proposed \krakenSpace language follows quite closely to John Shutt's definition of the pure Vau calculus (\cite{shutt2010fexprs}) augmented with explicit primitive data types and operations to create a more practical language.
Basing our language on the pure Vau calculus instead of the more complex calculi that form the basis of John Shutt's Kernel language makes the partial evaluation and optimization simpler and more feasible.
The main missing features from his complex calculi are mutation and continuations.
In order to keep our formalization understandable, we present a calculus simplified further from our real language that eliminates boolean values, varaidac parameters, and prunes down the builtin primitive combiners.

We start defining the \krakenSpace language by presenting its syntax, contexts, and small-step semantics before expounding upon the difference between its surface and internal syntax and semantics.

\subsection{Syntax}

\begin{figure}[h]
\[
  \begin{array}{rcll}
    n &\in& \mathbb{N}&\text{(Integers)}\\
    s &\in& Symbols&\\
    o &\in& \{\kprim{1}{eval}, \kprim{0}{vau},\kprim{1}{wrap}, \kprim{1}{unwrap},&\\
    &&\kprim{0}{if0}, \kprim{0}{vif0}, \kprim{1}{int-to-symbol},&\\
    &&\kprim{1}{symbol?}, \kprim{1}{int?}, \kprim{1}{combiner?},\kprim{1}{env?},&\\
    &&\kprim{1}{array?}, \kprim{1}{len}, \kprim{1}{idx}, \kprim{1}{concat},&\\
    &&\kprim{1}{+}, \kprim{1}{<=} \}&\text{(Primitive Operations)}\\
    E &:=& \kenv{(s \leftarrow T)\dots}{}{E} \alt \kenv{(s \leftarrow T)\dots}{s' \leftarrow E}{E} & \text{(Environments)}\\
    A &:=& (T \dots)& \text{(Arrays)}\\
    C &:=& \kcomb{n}{s'}{E}{(s\dots)}{T} & \text{(Combiners)}\\
    S &:=& n \alt o \alt E \alt C & \text{(Self evaluating terms)}\\
    V &:=& S \alt s \alt A & \text{(Values)}\\
    T &:=& V \alt AT & \text{(Terms)}\\
    AT &:=& \keval{T}{E} \alt \kcombine{T}{(T\dots)}{E} & \text{(Active terms)}\\
  \end{array}
  \]
  \caption{Syntax of the Base Language}
  \label{fig:basesyntax}
\end{figure}

The surface syntax of the language (Fig. \ref{fig:basesyntax}) consists of arrays of symbols and integers.
The internal syntax includes both primitive and derived combiners, environments, and actively-executing terms.
It is this split between surface syntax and internal syntax that permits a non-trivial theory \cite{shutt2010fexprs}.

\subsection{Contexts}
\begin{figure}[h]
\[
  \begin{array}{rcl}
    \Ctxt &:=& \square \alt \kcombine{\Ctxt}{(T\dots)}{E} \alt \kcombine{T}{(\Ctxt,T\dots)}{E}\\
    && \alt \kcombine{T}{(T\dots,\Ctxt,T\dots)}{E} \alt \kcombine{T}{(T\dots,\Ctxt)}{E}\\
  \end{array}
\]
  \caption{Contexts of the Base Language}
  \label{fig:basecontexts}
\end{figure}

The contexts for evaluation in our base calculus (Fig. \ref{fig:basecontexts}) define the holes where current evaluation must take place - namely, either evaluating the head of an array to find the combiner to call, or evaluating parameters before calling combiner.

\subsection{Small-Step Semantics}
\begin{figure}[h]
\[
  \begin{array}{rcl}
    \InCtxt{E}    &\rightarrow& \InCtxt{E'} ~ (\text{if } E \rightarrow E')\\
    \keval{S}{E}  &\rightarrow& S\\
    \keval{s}{E}  &\rightarrow& lookup(s,E)\\
    \keval{(T_1~T_2\dots)}{E}  &\rightarrow& \kcombine{\keval{T_1}{E}}{(T_2\dots)}{E}\\
    \\
    \kcombine{\kcomb{(S~n)}{s'}{E'}{(s\dots)}{Tb}}{(V\dots)}{E} &\rightarrow& \kcombine{\kcomb{n}{s'}{E'}{s}{Tb}}{\\&&\keval{V}{E}\dots}{E}\\
    \kcombine{\kcomb{0}{s'}{E'}{(s\dots)}{Tb}}{(V\dots)}{E} &\rightarrow& \keval{Tb}{\kenv{(s \leftarrow V)\dots}{s' \leftarrow E}{E'}}\\
    \\
    \kcombine{\kprim{(S~n)}{o}}{(V\dots)}{E} &\rightarrow& \kcombine{\kprim{n}{o}}{(\keval{V}{E}\dots)}{E}\\
  \end{array}
\]
  \caption{Semantics of the Base Language}
  \label{fig:basesemantics}
\end{figure}

Our small-step semantics (Fig. \ref{fig:basesemantics}) provides the evaluation steps for calls, symbols, and values. \krakenSpace evaluates self-evaluating values to themselves and symbols to the lookup of the symbol within the current environment.
The most interesting case is calls because it requires the first element of the array to be evaluated first resulting in a combiner. 

How the call proceeds after obtaining the combiner to be called depends on the "wrap level" and type of combiner.
This "wrap level" designates how many times the arguments to the combiner should be evaluated, and the semantics encodes evaluating the arguments that many times, enforcing that a combiner will have a "wrap level" of 0 (decremented every time the parameters are evaluated) before being called.
As we mentioned earlier, a combiner with a wrap level of 1 is an applicative, equivalent to functions in other languages.
The combiner can be either primitive (built into the language) or derived (created by the user with a call to \textit{vau}).
A derived combiner's body is evaluated in a new environment that maps the parameter symbols to the actual arguments, including mapping a special symbol to the current dynamic call environment, and chains up to the existing static environment stored within the derived combiner. The primitive combiners are explained more in the next section.

\subsection{Small-Step Semantics (selected primitives)}
\begin{figure}[h]
\[
  \begin{array}{rcl}
    \kcombine{\kprim{0}{eval}}{(V~E')}{E} &\rightarrow& \keval{V}{E'}\\
    \kcombine{\kprim{0}{vau}}{(s'~(s\dots)~V)}{E} &\rightarrow& \kcomb{0}{s'}{E}{(s\dots)}{V}\\
    \kcombine{\kprim{0}{wrap}}{\kcomb{0}{s'}{E'}{(s\dots)}{V}}{E} &\rightarrow& \kcomb{1}{s'}{E'}{(s\dots)}{V}\\
    \kcombine{\kprim{1}{unwrap}}{\kcomb{1}{s'}{E'}{(s\dots)}{V}}{E} &\rightarrow& \kcomb{0}{s'}{E'}{(s\dots)}{V}\\
    \kcombine{\kprim{0}{if0}}{(V_c~V_t~V_e)}{E} &\rightarrow& \kcombine{\kprim{0}{vif0}}{\\&&(\keval{V_c}{E}~V_t~V_e)}{E}\\
    \kcombine{\kprim{0}{vif0}}{(0~V_t~V_e)}{E} &\rightarrow& \keval{V_t}{E}\\
    \kcombine{\kprim{0}{vif0}}{(n~V_t~V_e)}{E} &\rightarrow& \keval{V_e}{E} ~\text{(n != 0)}\\
    \kcombine{\kprim{0}{int-to-symbol}}{(n)}{E} &\rightarrow& 'sn ~\text{(symbol made out of the number n)}\\
    \kcombine{\kprim{0}{array}}{(V\dots)}{E} &\rightarrow& (V\dots)\\
  \end{array}
\]
  \caption{Semantics of Base Language Primitives}
  \label{fig:baseprimitives}
\end{figure}
The semantics of the most important primitive combiners are given in Fg. \ref{fig:baseprimitives}, but are additionally described here in English for clarity:
\begin{itemize}
    \item $\kprim{0}{eval}$: evaluates its argument in the given environment.
    \item $\kprim{0}{vau}$: creates a new combiner and is analogous to lambda in other languages, but with a "wrap level" of 0, meaning the created combiner does not evaluate its arguments.
    \item $\kprim{0}{wrap}$: increments the wrap level of its argument. Specifically, we are "wrapping" a "wrap level" n combiner (possibly "wrap level" 0, created by \textit{vau}) to create a "wrap level" n+1 combiner. A wrap level 1 combiner is analogous to regular functions in other languages.
    \item $\kprim{0}{unwrap}$: decrements the "wrap level" of the passed combiner, the inverse of \textit{wrap}.
    \item $\kprim{0}{if}$: evaluates only its condition and converts to the $\kprim{0}{vif}$ primitive for the next step. It cannot evaluate both branches due to the risk of non-termination.
    \item $\kprim{0}{vif}$: evaluates and returns one of the two branches based on if the condition is non-zero. 
    \item $\kprim{0}{int-to-symbol}$: creates a symbol out of an integer.
    \item $\kprim{0}{array}$: returns an array made out of its parameter list.
\end{itemize}

The less interesting primitives we just describe here:
\begin{itemize}
  \item $\kcombine{\kprim{0}{type-test?}}{(A)}{E}$: \textit{array?}, \textit{comb?}, \textit{int?}, and \textit{symbol?}, each return 0 if the single argument is of that type, otherwise they return 1.
    \item $\kcombine{\kprim{0}{len}}{(A)}{E}$: returns the length of the single array argument.
    \item $\kcombine{\kprim{0}{idx}}{(A~n)}{E}$: returns the nth item array A.
    \item $\kcombine{\kprim{0}{concat}}{(A~B)}{E}$: combines both array arguments into a single concatenated array.
    \item $\kcombine{\kprim{0}{+}}{(A~A)}{E}$: adds its arguments
    \item $\kcombine{\kprim{0}{<=}}{(A~A)}{E}$: returns 0 if its arguments are in increasing order, and 1 otherwise.
\end{itemize}

\subsection{Base Language Summary}
This base calculus defined above is not only capable of normal lambda-calculus computations with primitives and derived user applicatives, but also supports a superset of macro-like behaviors via its support for operatives.
All of the advantages listed in the introduction apply to this calculus, as do the performance drawbacks, at least if implemented naively. Our partial evaluation and compilation framework will demonstrate how to compile this base language into reasonably performant binaries (WebAssembly bytecode, for our prototype).

\section{Partial Evaluation}
\label{sec:partial}

Partial Evaluation dates back at least to Lombardi's partial evaluator for Lisp \cite{lombardi1964lisp}.
In this work, we devise a partial evaluator specifically focused on partially evaluating away fexprs that behave like macros that we call \textbf{macro-like operatives}. By partially evaluating and inlining calls to these fexprs in a way analogous to macro expansion, we show a language based on fexprs instead of macros can be approximately as efficient and at least as powerful.
Our partial evaluation algorithm can eliminate all uses of macro-like operatives, where macro-like means the uses are static, and the combiner definition looks something like Listing \ref{code:macrolike}.
\begin{lstlisting}[language=Lisp,caption={A Macro-like Combiner Definition},label=code:macrolike]
(let (helper (lambda (...) <generate code like macro>))
     (vau dynamic_env (parameters) (eval (helper parameters) dynamic_env)))
\end{lstlisting}

In these cases, the combiner takes in its arguments unevaluated (as it is an operative), passes them to a helper function that generates new code based on the passed code parameters, and then immediately evaluates the code generated by the macro-like function.
Any combiner written following this template will be partially evaluated away by our algorithm.
Our partial evaluator and compiler can handle more cases than just this exact "macro-like" template, but we focus on this case.
By showing that it is possible to port anything written as a macro (which will be expanded away) to an analogous operative combiner (that will be partially evaluated away), we establish the practicality of using fexprs instead of macros as the fundamental building blocks of a purely functional Lisp-like language. When creating a new language, one loses neither expressivity nor significant performance by choosing fexprs as the base construct over the combination of functions and macros.
Any use of fexprs that is not partially evaluated away is something that \textit{could not be expressed via macros}, and thus paying a performance penalty in these cases is not onerous.

We implemented an online partial evaluation strategy for \kraken. Based on \cite{10.1145/243439.243447}'s description, offline partial evaluation relies on a Binding Time Analysis to have been done so it can determine which pieces of code are dependent upon dynamic runtime values of the system and which can be computed statically.
On the other hand, online partial evaluation actually executes the program using real values if they are statically known, and symbolic values otherwise.
Computations that use only static values result in more statically known values, whereas the computations involving dynamic values generate residual code to be included in the final program.

In our case, it is impossible to even determine what symbols will be used as variables without performing at least some execution, much less determining if those variables will contain static or dynamic values, so online partial evaluation is the solution for us.
By being an online partial evaluation algorithm, we do not mean it is happening at program run-time rather it runs at compile-time. In other words,
"online" only refers to when the binding analysis is done relative to partial evaluation, not when the partial evaluation itself is done. 

Sophisticated partial evaluation algorithms can handle partially-static data structures, in which only some of the data and structure is known \cite{10.1145/249069.231419}.
A classic example of partially-static data is a Lisp cons cell, one side of which contains a static integer and the other residual code.
Some advanced partial evaluators maintain additional information about values and residual code, such as its type \cite{ruf1993topics}.

The following is the key difficulty in compiling away macro-esque operatives - the call site environment must be reified and passed to the operative, and that reified environment will be a partially-static data structure.
That is, each frame in the environment is either static data, mapping symbols to static values, or it is a static description of dynamic data, containing symbols and the ID of the combiner that created it (by being called).
Either of these types of frames may then chain upwards to a parent frame, which may be of either type as well.
Correctly handling these partially-static-data environments while ensuring that all macro-esque operatives are partially evaluated away while preventing both non-termination and exponential runtime was the key challenge in this work.
A thorny complication is that that a combiner definition may have to be partially evaluated multiple times in environments with different amounts of static data before it has enough information to reduce as far as it should according to our criteria of eliminating all statically known operative calls.
This means if partial evaluation for a form fails, we can't turn it into residual code right away, as it might be evaluated again later.
If this is done naively, it is very easy to incur exponential runtime as at every call first the combiner is evaluated, then the parameters, followed by the combiner again.
Since this compounds at every call inside a combiner (and compounds via environments containing combiners containing environments without memorization or similar), a mitigating technique is needed.
Our solution to this issue is the needed-for-progress-IDs system, which will be explained below.

As bits of the partial evaluation, especially as relating to calls, get complex, we broke down the algorithm into sections to make it easier to understand. In each section, we provide examples along the way, and at the end provide some final examples to show the process. Our partial evaluation roadmap:
\begin{itemize}
    \item \textbf{Section 4.1}: Partial Evaluation Contexts, Marking Syntax, and Unval Relation
    \item \textbf{Section 4.2}: Small Step Semantics
    \item \textbf{Section 4.3}: Helper Relations
    \item \textbf{Section 4.4}: Partial-Eval Versions of Primitives
    \item \textbf{Section 4.5}: Total Effect of Partial Evaluation
    \item \textbf{Section 4.6}: Examples
\end{itemize}

\subsection{Evaluation Contexts, Marking Syntax, and Unval Relation}
For partial evaluation, we added a new active term, \textit{under}, replace \textit{eval} with \textit{peval} (partial eval), and add new values to \textit{combine}, as shown in Fig. \ref{fig:extendedtermsyntax}.
\begin{figure}[h]
\[
  \begin{array}{rccll}
    AT &:=&& \kpval{T}{\kmkd{i_x}{E}}{ES}{FS}\\
       && \alt & \kunder{T}{(T\dots)}{\kmkd{i_x}{E}}{ES}{FS}\\
       && \alt & \kcombine{T}{(T\dots)}{E} & \text{(Active terms)}
  \end{array}
  \]
  \caption{New Terms for Partial Evaluation}
  \label{fig:extendedtermsyntax}
\end{figure}

We track two critical types of information throughout the partial evaluation process: IDs of environments that contain values needed for progress, and a set of the currently evaluating terms.
This allows us to massively prune the execution space and guide the path of partial evaluation.

This bookkeeping information is added via a "mark" pass.
Partial evaluation will operate on this "mark"'d representation, as will the compiler.
The extra information that bookkeeping adds for each type of form is as follows:
\begin{itemize}
    \item \textbf{Combiner}: A unique ID ($i$) that indicates environments created by calling this combiner\\
    Example: $\kmkd{i}{\kcomb{n}{s'}{\kmkd{i'_{f'}}{E'}}{(s\dots)}{Tb}}$
    \item \textbf{Env}: A unique ID ($i_r$ or $i_f$) matching this environment to the combiner whose call created it. The $r$ or $f$ subscript indicating whether the environment is "fake", a static description of the dynamic environment which maps symbols to placeholder values, or "real", a fully static map from symbols to (almost entirely) static values. "Almost entirely" because a "real" environment can map a symbol to a "fake" environment, as happens when partially evaluating a call to an operative that takes in its call site environment. \\
    Example: $\kmkd{i_r}{\kenv{(s \leftarrow V)\dots}{s' \leftarrow \kmkd{i_x}{E}}{\kmkd{i'_{x'}}{E'}}}$\\
    Example: $\kmkd{i_f}{\kenv{(s \leftarrow V)\dots}{s' \leftarrow \kmkd{i_x}{E}}{\kmkd{i'_{x'}}{E'}}}$
    \item \textbf{Symbol}: ID of environment that the symbol will be resolved in, if applicable. True means the symbol has yet to be partially evaluated, so the ID of the environment will resolve to unknown. $\emptyset$ means the symbol is a value instead of a suspended lookup.\\
    Example: $\kmkd{\truev}{s}$\\
    Example: $\kmkd{7}{s}$\\
    Example: $\kmkd{\emptyset}{s}$
    \item \textbf{Array}: Arrays can be marked by one of three options: \textit{val}, \textit{freshCall}, or \textit{attemptedCall}. \textit{val} indicates the array is a value. \textit{freshCall} indicates the array is a call that hasn't been partially evaluated yet.  \textit{attemptedCall} indicates the suspended call was partially evaluated, but couldn't proceed. \textit{attemptedCall} contains two values - the ID of the dynamic calling environment if the called combiner needs it (otherwise $\emptyset$) and/or the (form, environment) pair that caught and prevented infinite recursion.\\
    Example: $\kmkd{\kval}{(V_1~V_2)}$\\
    Example: $\kmkd{\kfresh}{(V_1~V_2)}$\\
    Example: $\kmkd{\katt{\emptyset}{\emptyset}}{(V_1~V_2)}$
\end{itemize}

In our implementation, we store additional data for efficiency, such as all for-progress-IDs found inside an array, but we will define them in Appendix A as stand-alone functions for ease of presentation.
Our implementation essentially pre-computes and stores the results of these functions as additional bookmarking data, reducing exponential lookups to constant factors.
This could also be achieved via memoization.

The mark relation in Fig. \ref{fig:markrelation} only needs to add bookkeeping to the surface syntax, so it is quite simple.
\begin{figure}[h]
\[
  \begin{array}{rcll}
    \kmark{n} &=& n\\
    \kmark{s} &=& \kmkd{\emptyset}{s}\\
    \kmark{(T\dots)} &=& \kmkd{\kval}{(\kmark{T}\dots)}\\
  \end{array}
  \]
  \caption{Mark Relation}
  \label{fig:markrelation}
\end{figure}

We must also split evaluation into two pieces: "unval"-ing and "partial-eval"-ing.
Unvaling takes a value and turns it into a suspended computation. Partial-evaling takes a suspended computation (or a value containing one) and tries to perform at least some evaluation in order to reduce work at runtime.
Now, evaluation is just unvaling composed with partial evaluation.

\begin{figure}[h]
\[
  \begin{array}{rcll}
    \kunval{n} &=& n\\
    \kunval{\kprim{n}{o}} &=& \kprim{n}{o}\\
    \kunval{\kmkd{i''}{\kcomb{n}{s'}{\kmkd{i'_{f'}}{E'}}{(s\dots)}{Tb}}} &=& \kmkd{i''}{\kcomb{n}{s'}{\kmkd{i'_{f'}}{E'}}{(s\dots)}{Tb}}\\
    \kunval{\kmkd{i}{E}} &=& \kmkd{i}{E}\\
    \kunval{\kmkd{\emptyset}{s}} &=& \kmkd{\truev}{s}\\
    \kunval{\kmkd{\kval}{(T_1~T_2\dots~T_n)}} &=& \kmkd{\kfresh}{(\kunval{T_1}~T_2\dots~T_n)}\\
  \end{array}
  \]
  \caption{Unval Relation}
  \label{fig:unvalrelation}
\end{figure}
Unvaling (Fig. \ref{fig:unvalrelation}) a self-evaluating value is just the value itself, as there is no computation to suspend.
The two forms that change when unvaling are symbols (which go from values to suspended lookups of that symbol in an environment) and arrays (which turn into suspended calls).

Contexts in Fig. \ref{fig:pecontexts} are defined like before, but augmented with our new "peval" (partial evaluate), combiner bodies, and multiple positions in the new "under" form.
\begin{figure}[h]
\[
  \begin{array}{rcl}
    \Ctxt &:=& \square \alt \kpval{\Ctxt}{\kmkd{i_x}{E}}{ES}{FS} \alt \kcomb{n}{s'}{\kmkd{i_x}{E}}{(s\dots)}{\Ctxt}\\
    && \alt \kunder{\Ctxt}{(T\dots)}{\kmkd{i_x}{E}}{ES}{FS} \alt \kunder{T}{(\Ctxt~T\dots)}{\kmkd{i_x}{E}}{ES}{FS}\\
    && \alt \kunder{T}{(T~\Ctxt~T\dots)}{\kmkd{i_x}{E}}{ES}{FS} \alt \kunder{T}{(T\dots~\Ctxt)}{\kmkd{i_x}{E}}{ES}{FS}\\
    && \alt \kcombine{\Ctxt}{(T\dots)}{E} \alt \kcombine{T}{(\Ctxt,T\dots)}{E}\\
    && \alt \kcombine{T}{(T\dots,\Ctxt,T\dots)}{E} \alt \kcombine{T}{(T\dots,\Ctxt)}{E}\\
  \end{array}
\]
  \caption{Contexts for Partial Evaluation}
  \label{fig:pecontexts}
\end{figure}

\subsection{Partial-Eval Small-Step Semantics}
In Appendix B we present some simplified pseudocode to give an additional reference and roadmap for the relations presented here.

For simplicity, we split the semantics into two subsections. The first one will talk about partial evaluation of everything but calls. The second one will just be the calls due to their complexity. 

\subsubsection{Partial Evaluation of Non-Calls}: The partial evaluation of non-calls is easier to understand, and it helps to get a feel for the basic semantics of partial evaluation.

\begin{figure}[h]
\[
  \begin{array}{rcl}
    \InCtxt{E}    &\rightarrow& \InCtxt{E'} ~ (\text{if } E \rightarrow E')\\
    \\
    \kpval{x}{\kmkd{i_x}{E}}{ES}{FS}  &\rightarrow& x~~\text{if}~\truev~\notin~\kprog{x}~\land\\
                                                  &&~\kprog{x}~\cap~ES~=~\emptyset~\land\\
                                                  &&~\kiprog{x}~\cap~ES~=~\kiprog{x}\\
                                                  && \textbf{else:}\text{ continue below}\\
                                                  \\
    \kpval{\kmkd{x}{s}}{\kmkd{i_x}{E}}{ES}{FS}  &\rightarrow& lookup(s,E)\\
    \kpval{\kmkd{i'_{x'}}{E'}}{\kmkd{i_{x}}{E}}{ES}{FS}  &\rightarrow& \kmkd{i'_{x'}}{E''}~\textbf{if}~\kmkd{i'_{x''}}{E''} \in ES ~\textbf{else}~\kmkd{i'_{x'}}{E'}\\
    \\

    \kpval{&&\\ \kmkd{i''}{\kcomb{n}{s'}{\kmkd{i'_r}{E'}}{(s\dots)}{Tb}}}{&&\\\kmkd{i_x}{E}}{ES}{FS}  &\rightarrow& \kmkd{i''}{\kcomb{n}{s'}{\kmkd{i'_r}{E'}}{(s\dots)}{Tb}}\\
    \\
    \kpval{&&\\ \kmkd{i''}{\kcomb{n}{s'}{\kmkd{i'_{f'}}{E'}}{(s\dots)}{Tb}}}{&&\\\kmkd{i_x}{E}}{ES}{FS}  &\rightarrow& \text{let}~E''=\kenv{(s \leftarrow \kmkd{i''}{s})\dots}{s' \leftarrow \kmkd{i''}{s'}}{\kmkd{i_x}{E}}~\text{in}\\
    && \kmkd{i''}{\kcomb{n}{s'}{\kmkd{i_x}{E}}{(s\dots)}{\\&&\qquad \kpval{Tb}{\kmkd{i''_f}{E''}}{\{\kmkd{i''_f}{E''}\}\cup ES}{FS}}}\\
  \end{array}
\]
  \caption{Non-Call Partial-Eval Semantics}
  \label{fig:ncpesemantics}
\end{figure}
In Fig. \ref{fig:ncpesemantics}, we see the formal relations for partial evaluation of all forms except calls. 
The components of an actively partially evaluating form $\kpval{x}{\kmkd{i_x}{E}}{ES}{FS}$ is as follows:
\begin{itemize}
    \item The form that needs to be partially evaluated ($x$ in the figure).
    \item The environment to partially evaluate it in ($\kmkd{i_x}{E}$ in the figure)
    \item The "call stack" (a set in this case) of environments for evaluating the forms above this one ($ES$ in the figure)
    \item A set of currently evaluating forms ($FS$ in the figure) used to check for and prevent infinite recursion
\end{itemize}


Our first relation is the same as from the small-step semantics of the base language - that a larger expression can step if a sub-component of it as indicated by the Context form can step.

The next relation checks to see if partially evaluating the form will make any progress.
There are three conditions which, if true, mean that the form cannot make progress:
\begin{itemize}
    \item the form has already been partially evaluated ($\truev~\notin~\kprog{x}$)
    \item none of the environments the form needs to progress are currently real or in our environment stack ($\kprog{x}~\cap~ES~=~\emptyset$)
    \item the form hadn't previously stopped evaluating to prevent infinite recursion or we are no longer under the form that began the loop ($\kiprog{x}~\cap~ES~=~\kiprog{x}$)
\end{itemize}
If all three of these are true, partially evaluating the form won't have any affect and instead the form can be returned immediately.

On the other hand, if any of these are false the form might be able to make progress and it falls through to the remaining relations.
Note that integers, symbol values, and array values will all already have been returned by the progress-check condition.
We are only left with suspended symbol lookup, environment value, or derived combiner.
For suspended symbol lookup, we just lookup the symbol in the current environment. If this environment is fake, it will return the same symbol marked with the ID of the environment that it would be found in if it were real.
When partially evaluating an environment value we check the current environment\_stack to see if the environment has a newer, real version (identified by ID) - if so we return it, else we return the environment unchanged.

A derived combiner is the most complicated case in this section.
First, we check to see if its static environment is real or not.
If it is, the fifth rule of Fig. \ref{fig:ncpesemantics} (as noted by the r subscript in the static environment $\kmkd{i'_r}{E'}$ in the combiner value), we return it immediately, since this combiner has already been evaluated in a real environment.
As a closure, it captured the environment from its original evaluation during its creation, and since it might have since moved this environment should not be replaced.
On the other hand, a fake static environment (as noted by the f subscript in the static environment $\kmkd{i'_f}{E'}$ in the combiner value) means the closure's initial evaluation has not yet happened and the closure is still at its original defining place.
In this case, we replace the old static environment with the current one, and then create a new fake environment that maps the function's parameters to placeholder symbols marked with the ID of the combiner/fake environment. We further partially evaluate the combiner's body in this fake environment.
The final result is a new derived combiner form that wraps up the new static environment and the partially evaluated body form.

\subsubsection{Partial Evaluation of Calls}
The partial evaluation of calls is more complex due to the high amount of bookkeeping that needs to be maintained. This complexity can be seen in Fig. ~\ref{fig:cpesemantics}.

\begin{figure}[h]
\[
  \begin{array}{rcl}
    \kpval{\kmkd{\kfresh}{(T_1~T_2\dots)}}{\kmkd{i_x}{E}}{ES}{FS}   &\rightarrow& \kpcombine{\\&&\qquad\kpval{T_1}{\kmkd{i_x}{E}}{ES}{FS}}{\\&&\quad(T_2\dots)}{\kmkd{i_x}{E}}{ES}{FS}\\
    \kpval{\kmkd{\katt{x}{y}}{(T_1~T_2\dots)}}{\kmkd{i_x}{E}}{ES}{FS}   &\rightarrow& \kpcombine{\\&&\qquad\kpval{T_1}{\kmkd{i_x}{E}}{ES}{FS}}{\\&&\quad(T_2\dots)}{\kmkd{i_x}{E}}{ES}{FS}\\
    \\
    \kpcombine{\kmkd{x}{s}}{(T_2\dots)}{\kmkd{i_x}{E}}{ES}{FS} &\rightarrow& \kmkd{\katt{\emptyset}{\emptyset}}{(\kmkd{x}{s}~T_2\dots)}\\
    \kpcombine{\kmkd{\katt{x}{y}}{(T_1\dots)}}{&&\\(T_2\dots)}{\kmkd{i_x}{E}}{ES}{FS} &\rightarrow& \kmkd{\katt{\emptyset}{\emptyset}}{\\&&\qquad(\kmkd{\katt{x}{y}}{(T_1\dots)}~T_2\dots)}\\
    \\
    \kpcombine{\kprim{(S~n)}{o}}{(V\dots)}{\kmkd{i_x}{E}}{ES}{FS} &\rightarrow& \kpcombine{\kprim{n}{o}}{\\&&\qquad\kpval{\kunval{\kpval{V}{\kmkd{i_x}{E}}{ES}{FS}}}{\\&&\qquad\qquad\kmkd{i_x}{E}}{ES}{FS}\dots}{\kmkd{i_x}{E}}{ES}{FS}\\
    \\
    \kpcombine{&&\\ \kmkd{i''}{\kcomb{(S~n)}{s'}{\kmkd{i'_{x'}}{E'}}{(s\dots)}{Tb}}}{&&\\ (V\dots)}{\kmkd{i_x}{E}}{ES}{FS} &\rightarrow& \kpcombine{\kmkd{i''}{\kcomb{n}{s'}{\kmkd{i'_{x'}}{E'}}{s}{Tb}}}{\\&&\qquad\kpval{\kunval{\kpval{V}{\kmkd{i_x}{E}}{ES}{FS}}}{\\&&\qquad\qquad\kmkd{i_x}{E}}{ES}{FS}\dots}{\kmkd{i_x}{E}}{ES}{FS}\\
    \\
    \kpcombine{&&\\ \kmkd{i''}{\kcomb{0}{s'}{\kmkd{i'_{x'}}{E'}}{(s\dots)}{Tb}}}{&&\\(V\dots)}{\kmkd{i_x}{E}}{ES}{FS} &\rightarrow&  \textbf{let}~E''=\kenv{(s \leftarrow V)\dots}{s' \leftarrow \kmkd{i_x}{E}}{\kmkd{i'_{x'}}{E'}}~\textbf{in}\\
    &&\textbf{let}~C = \kmkd{i''}{\kcomb{0}{s'}{\kmkd{i'_{x'}}{E'}}{(s\dots)}{Tb}}~\textbf{in}\\
    &&\textbf{let}~F = (C,E'')~\textbf{in}\\
    &&\kmkd{\katt{\emptyset}{F}}{(C~V\dots)}~\textbf{if}~F \in FS\\
    &&\textbf{else}~\kunder{\\
    &&\kpval{Tb}{\kmkd{i''_r}{E''}}{(\kmkd{i''_r}{E''}\cup ES)}{\{F\}\cup FS}}{\\&&(\kmkd{i''}{\kcomb{0}{s'}{\kmkd{i'_{x'}}{E'}}{(s\dots)}{Tb}}~V\dots)}{\\&&\kmkd{i_x}{E}}{ES}{FS}\\
    \\
    \kunder{z}{&&\\(\kmkd{i''}{\kcomb{0}{s'}{\kmkd{i'_{x'}}{E'}}{(s\dots)}{Tb}~V\dots)}}{&&\\\kmkd{i_x}{E}}{ES}{FS} &\rightarrow& \textbf{let}~o = returnOk(z,i')~\textbf{in}\\
    &&dropRV(z,\kmkd{i_x}{E},ES,FS)~\textbf{if}~o=\truev~\textbf{else}\\
    &&\textbf{let}~i_?= i_x~\textbf{if}~s' \neq \emptyset~\textbf{else}~\emptyset~\textbf{in}\\
    &&\kmkd{\katt{i_?}{\emptyset}}{(\kmkd{i''}{\\&&\qquad\kcomb{0}{s'}{\kmkd{i'_{x'}}{E'}}{(s\dots)}{Tb}}V\dots)}\\
  \end{array}
\]
  \caption{Call Partial-Eval Semantics}
  \label{fig:cpesemantics}
\end{figure}
First, we partially evaluate the first item in the array, which will be the combiner, transitioning to a \textit{combine} form to indicate we are in the process of evaluating a call, as seen before in the base language semantics in Section 3.
The next rule handles the case where the thing to be called isn't a combiner but instead a suspended symbol lookup ($\kmkd{x}{s}$). In this case, we simply return the call as-is, marked as attempted, because we can't make progress.
We do the same if the thing to be called is another suspended call ($\kmkd{\katt{x}{y}}{(T_1\dots)}$).
Otherwise, we have a combiner, either primitive or derived, and we extract the wrap\_level from it. We then partial eval, unval, and partial eval again the arguments wrap\_level number of times.
If the combiner is a normal applicative, this would mean 1 time, like function calls in most other languages. The other common option is 0 times when the combiner is an operative, in order to take the code for the parameters in unevaluated. This might be so the operative can behave like a macro, or implement special control flow.
Numbers greater than 1 for wrap\_level are also possible, but rarely useful.
We retain the ability for wrap\_levels greater than 1 for uniformity, but have not used them in practice.

Now, we are ready to execute the actual call due to the parameters being evaluated the proper number of times according to the wrap\_level in the combiner.
The semantics for executing calls to primitive combiners are given as relations in Fig. 9 in Appendix A.
For derived combiners, we create a new inner environment ($E''$) that maps the parameter symbols to the argument values, the special dynamic environment parameter symbol ($s'$) to the current dynamic environment ($\kmkd{i_x}{E}$), and chains up to the indicated static environment ($\kmkd{i'_{x'}}{E'}$) from the combiner value.
We first check to see if this (form, environment) pair is currently executing ($(C,E'') \in FS$), and return early if so to prevent infinite recursion.
If it does return early, the returned call notes the (form, environment) pair that caused it to stop executing in its \textit{atmdCall} form: $\kmkd{\katt{\emptyset}{F}}{(C~V\dots)}$.
Otherwise, the combiner body is evaluated in this new environment, with the call site marked by the \textit{under} marker form.
The \textit{under} form delineates where a currently executing call is. This allows us to check whether the value/code resulting from partially evaluating a call site is legal to return whenever the form stops evaluating.
To do this, it contains both the partial evaluation of the combiner body to execute the call as well as a fallback term.
The fallback term, which is an updated suspended call in this case, is to be used if the call fails.
The fallback term additionally carries the current dynamic environment, environment stack, and executing form set to be used in the next rule.
The result is checked by the \textit{returnOk} auxiliary function to see if this value is safe to return (defined in Appendix A, overview given in Section 4.3). If it is safe, it passes through \textit{dropRV} before being returned (Fig. 8 in Appendix A).

The case where it is not legal to return a value is if there is, or could be, a reference that must be resolved in the inner environment created by this call.
If the result is legal to return, the dropRedundentVeval (or \textit{dropRV}) ensures no useless call to \textit{veval} (a modified \textit{eval} used in the primitive semantics (Fig. 9 in Appendix A)) to wrap any part of the result.
This is important because these unnecessary wrapper calls to \textit{veval} can block partial evaluation progress in some cases.

When the result form is not legal to return, we return a call form with the partially evaluated combiner (with updated wrap\_level) and partially evaluated arguments based on the fallback value in the \textit{under} form.
If the combiner does take in a dynamic environment, we note the current dynamic environment's ID on the suspended call form ($i_?$).

\subsection{Helper Relations}
Next we quickly describe the various helper relations (fully defined in Appendix A) used in our definition of the partial evaluation relations.
\begin{itemize}
    \item \textbf{Needed-For-Progress}: The needed-for-progress relation shows the set of IDs for which at least one needs to correspond to a real environment (static data) with values in order for the form to make progress. $\truev$ means that it can make progress no matter what.
    \item \textbf{Needed-For-Progress-Upper}: An "upper" version of the needed-for-progress relation (in Appendix A) that tracks IDs of environments that are real themselves, but chain upwards to fake environment IDs.
    \item \textbf{Needed-For-Progress-Infinite}: This relation extracts the forms that have previously stopped executing to prevent infinite recursion. This is important when we are not currently executing in the call stack of one of these forms. If we are not, we should keep executing this form, as it has more evaluation to go before it hits another opportunity for infinite recursion. This is part of the mechanism that allows us to use the Y-combinator to implement recursion without either having infinite recursion or un-evaluated recursive calls with finite inputs.
    \item \textbf{Lookup}: As its names signifies, it finds the value associated with a symbol in an environment.
    \item \textbf{returnOk}: It determines if it is legal to return a particular result out of a particular environment. For example, a term can be returned out of a call if the ID of the inner environment/combiner is not present in the term either explicitly or implicitly through a suspended call to a combiner that takes in its dynamic environment.
    \item \textbf{IDin}: It determines if this ID appears in this value without being under a combiner that introduces this ID.
    \item \textbf{takesDE}: It determines if this combiner takes in the dynamic environment.
    \item \textbf{dropRV} / dropRedundentVeval: Our final helper function removes extraneous calls to \textit{veval} that can get in the way of further partial evaluation.
    \textit{veval} is a version of the applicative \textit{eval}, but with both parameters (the term to evaluate and the environment to evaluate it in) already unvaled and partially evaluated.
    It requires special handling (which it receives via the -1 wrap\_level), because its term argument should not be partially evaluated via the normal machinery, which would use the wrong environment (the current dynamic environment, instead of the explicit environment passed to \textit{veval}/\textit{eval}).
Calls to \textit{eval} are common in macro-like operatives, where the normal final call of a macro-like operative is to \textit{eval} the code constructed during the body of the combiner.
This call to \textit{eval} will partially evaluate to a call to \textit{veval}, which will then return successfully to the call site.
In a normal macro-like operative call, this call site's dynamic environment is the explicit environment passed to \textit{veval}, and thus the call to \textit{veval} is extraneous and the term will be inlined directly, completing the partial evaluation dance that fully expands macro-like operative calls.
This removal is the responsibility of \textit{dropRV}.
A "macro-like operative combiner call" example that demonstrates this happening is located below in section 4.6.
\end{itemize}

\subsection{Partial Eval Primitives Small-Step Semantics}
Finally, we come to the (selected) semantics of the primitive combiners.
For space, their formal definition is relegated to Appendix A. In general, they perform the same function as the simpler base primitives, but operate on the marked terms.
The main combiners of interest are \textit{if0} and \textit{vau} which perform additional partial evaluation on their branches and body, respectively.
We have already heard about \textit{eval}'s half-life as \textit{veval} and function in carrying along the proper environment to evaluate a suspended piece of code in until it can be unwrapped by \textit{dropRV} and spliced into its final location.
The formal definition of \textit{eval}, \textit{veval}, and \textit{dropRV} are in Appendix A.
The primitive implementations omitted are the trivial ones - they only evaluate if their parameters are fully evaluated values and they evaluate to the same value that they would under the base semantics.

\subsection{Combined Effect of Partial Evaluation}
The end result of all of this interconnected machinery is the removal of all statically called operatives where the operative was written in a macro-esque style.
Note that the invariants maintained by partial evaluation ensure that derived combiners are only executed whose arguments are all values.
Additionally, the result of a call to a combiner is either a value, a suspended computation where all calls either don't take in the dynamic environment (which would be the combiner's environment that it's being returned from), or an explicit call to veval providing its own environment.
These are the cases that can be returned by a macro-like combiner!
In addition, the partial evaluation process strips redundant calls to veval.
Once the suspended code is returned from the call to the macro-like combiner, its call to veval will become redundant and the inner suspended computation will be inlined into the parent, just like how a macro would be expanded to code spliced into its calling location.

\subsection{Examples}
To illustrate the algorithm, we have some examples of marking, unvaling, and then partial-evaling in stages. The full step-by-step examples are too long for any paper, so these have been somewhat abbreviated.
\subsubsection{Addition Example}
We'll walk through the full partial evaluation steps for $(+~1~2)$ which is 3.
\begin{longtable}{cc}
      $(+~1~2)$ & The initial code\\
      $\kmkd{\kval}{(\kmkd{\emptyset}{+}~1~2)}$ & Marked\\
      $\kmkd{\kfresh}{(\kmkd{\truev}{+}~1~2)}$ & Then unvaled\\
      \\
      $\kpval{\kmkd{\kfresh}{(\kmkd{\truev}{+}~1~2)}}{\kmkd{i_r}{E}}{ES}{FS}$ & We can't show the entire env,\\
      &but for illustration say that E\\
      &maps "+" to the primitive + combiner\\
      \\
      $\kpcombine{\kpeval{\kmkd{\truev}{+}}{\kmkd{i_r}{E}}{ES}{FS}}{(1~2)}{\kmkd{i_r}{E}}{ES}{FS}$ & Begin call, PV combiner\\
      $\kpcombine{\kprim{1}{+}}{(1~2)}{\kmkd{i_r}{E}}{ES}{FS}$ & Lookup replaces the symbol +\\
      & with the primitive \\&combiner\\
      \\
      $\kpcombine{\kprim{0}{+}}{(\kpval{\kunval{1}}{\kmkd{i_r}{E}}{ES}{FS}$ & Unval+PartialEval to evaluate\\
      $\kpval{\kunval{2}}{\kmkd{i_r}{E}}{ES}{FS})}{\kmkd{i_r}{E}}{ES}{FS}$ & parameters, but integers stay the \\&same\\
      \\
      $\kpcombine{\kprim{0}{+}}{(1~2)}{\kmkd{i_r}{E}}{ES}{FS}$ & And then the call\\
      +(1~2) & Primitive does the calculation\\
      3 & result is 3, as expected,\\
        & which is legal to return\\
\end{longtable}
\subsubsection{Constant Combiner Example}
Now that we've done addition, we'll get slightly more complex by introducing the creation of a combiner with a body that can be partially evaluated. We take larger steps because we've gone over the details with the simple addition. In this case, we have $(vau~(x)~(+~1~2~x))$ meaning "1+2+x" for some input x.
We use just a pinch of syntactic sugar to have a 2-argument vau that is equivalent to the 3-argument vau that ignores its special dynamic environment parameter.
\begin{longtable}{cc}
      $(vau~(x)~(+~1~2~x))$ & The initial code\\
      $\kmkd{\kval}{(\kmkd{\emptyset}{vau}~\kmkd{\kval}{(x)}~\kmkd{\kval}{(\kmkd{\emptyset}{+}~1~2~x)})}$ & \text{Parsed and marked syntax}\\
    $\kmkd{\kfresh}{(\kmkd{\truev}{vau}~\kmkd{\kval}{(x)}~\kmkd{\kval}{(\kmkd{\emptyset}{+}~1~2~x)})}$ & \text{Unvaled}\\
      &\\
      $\kpval{\kmkd{\kfresh}{$&\text{We can't show the entire env.}\\$(\kmkd{\truev}{vau}~\kmkd{\kval}{(x)}~\kmkd{\kval}{(\kmkd{\emptyset}{+}~1~2~x)})$&\text{For illustration, E maps "vau"}\\$}}{\kmkd{i_r}{E}}{ES}{FS}$ &\text{to the primitive vau combiner}\\
    \\
      $\kpcombine{\kpeval{\kmkd{\truev}{vau}}{\kmkd{i_r}{E}}{ES}{FS}}{$&\\$(\kmkd{\kval}{(x)}~\kmkd{\kval}{(\kmkd{\emptyset}{+}~1~2~x)}))}{\kmkd{i_r}{E}}{ES}{FS}$ & \text{Begin call, PV combiner}\\
      &\\
      $\kpval{$&\text{The symbol Vau maps to its}\\
      $\kmkd{7}{\kcomb{0}{s'}{\kmkd{i_f}{E}}{(x)}{\kmkd{\kfresh}{(\kmkd{\truev}{+}~1~2~\kmkd{\emptyset}{x})}}}}{$ &\text{combiner value that will now}\\
      $\kmkd{i_r}{E}}{ES}{FS}$ &\text{be partially evaluated}\\
      \\
      $\kmkd{7}{\kcomb{0}{s'}{\kmkd{i_f}{E}}{(x)}{$&\text{Partial evaluating the body}\\$\kpval{\kmkd{\kfresh}{(\kmkd{\truev}{+}~1~2~\kmkd{\emptyset}{x})}}{$& \text{with fake environment. Notice, we}\\$\kmkd{7_f}{\kenv{(x \leftarrow \kmkd{7}{x})}{}{\kmkd{i_r}{E}}}}{ES}{FS}}}$ & \text{are almost back to our first example}\\
      &\\
      $\kmkd{7}{\kcomb{0}{s'}{\kmkd{i_r}{E}}{(x)}{\kmkd{\katt{\emptyset}{\emptyset}}{(\kprim{0}{+}~3~\kmkd{7}{x})}}}$ & \text{We'll fast forward through the}\\
      & \text{process from our first example}\\
\end{longtable}
We moved quickly through the part mostly shared with the previous example. The only difference being the addition of the parameter reference $x$ to the addition.
The only thing to note is that $\kmkd{\emptyset}{x}$ is unvaled to $\kmkd{\truev}{x}$ (not shown) and then was partially evaluated to $\kmkd{7}{x}$ (7 being the ID of the combiner/fake environment). Furthermore, the partial evaluation version of addition had to return a new partially evaluated call, since it could not yet evaluate $x$.
We are left with a partially-evaluated combiner, but haven't yet seen this new technique do anything that previous partial evaluation techniques couldn't.
For that, let's see a high-level view of how a macro-like operative call would be partially evaluated away.
\subsubsection{Macro-like Operative Call Example}
Due to the length of a step-by-step evaluation of this code, we will take larger jumps than before, omitting that which could be inferred from our two previous examples.
\begin{lstlisting}[language=Lisp,caption={},label=code:additionexample]
(let ( (double_parameter (vau de (x) (eval (array + x x) de))) )
     (vau (x) (double_parameter (+ 1 2 x))))
\end{lstlisting}
We'll skip over how "let" works for now (it's an operative combiner too) and focus on the partial evaluation of a macro-like operative part.
This piece of code defines a macro-like operative called "double\_parameter" that takes in a piece of code "x" unevaluated along with the dynamic calling environment "de" and then constructs the code "(+ x x)" as an array and evaluates it in de, the calling environment.
This code is intentionally simplistic and will evaluate its argument twice (though the downsides of doing so are considerably reduced in a purely functional language where evaluating x cannot have side effects).
This should bring to mind the classic macro example from C, "\#define double(x) (x+x)", and indeed we chose it for its familiarity.
Let's start by seeing what the macro-like f-expression itself looks like partially evaluated, and then we'll jump right into its application. This code:
\begin{lstlisting}[language=Lisp,caption={},label=code:additionexample]
(vau de (x) (eval (array + x x) de))
\end{lstlisting}
becomes
\[
      \kmkd{6}{\kcomb{0}{de}{\kmkd{i_r}{E}}{(x)}{\kmkd{\katt{\emptyset}{\emptyset}}{(\kprim{0}{eval}~\kmkd{\katt{\emptyset}{\emptyset}}{(\kprim{0}{array}~\kprim{1}{+}~\kmkd{6}{x}~\kmkd{6}{x})}~\kmkd{6}{de})}}}
\]
in a way very similar to our earlier examples of partially evaluated combiners.
We have a combiner with two suspended calls, nested, with some suspended symbol lookups.\\
Now let's look at its use, when partially evaluating the body:
\begin{lstlisting}[language=Lisp,caption={},label=code:additionexample]
(vau (x) (double_parameter (+ 1 2 x)))
\end{lstlisting}
becomes, skipping forwards to evaluating the body
\begin{longtable}{cc}
      $\kmkd{7}{\kcomb{0}{\emptyset}{\kmkd{i_f}{E}}{(x)}{\kpval{$&\\
      $\kmkd{\kfresh}{(\kmkd{\truev}{double\_parameter}~\kmkd{\kval}{(\kmkd{\emptyset}{+}~1~2~\kmkd{\emptyset}{x})})}}{$&\\
      $\kmkd{i_r}{\kenv{(x \leftarrow \kmkd{7}{x})}{}{\kmkd{i_r}{E}}}}{ES}{FS}}}$ & \text{forwards to call}\\
      \\
      $\kmkd{7}{\kcomb{0}{\emptyset}{\kmkd{i_f}{E}}{(x)}{\kpval{\kmkd{\kfresh}{($
      &\\
      $\kmkd{6}{\kcomb{0}{de}{\kmkd{i_r}{E}}{(x)}{\kmkd{\katt{\emptyset}{\emptyset}}{$
      &\\
      $(\kprim{0}{eval}~\kmkd{\katt{\emptyset}{\emptyset}}{(\kprim{0}{array}~\kprim{1}{+}~\kmkd{6}{x}~\kmkd{6}{x})}~\kmkd{6}{de})}}}$
      &\\
      $\kmkd{\kval}{(\kmkd{\emptyset}{+}~1~2~\kmkd{\emptyset}{x})})}}{$&\\
      $\kmkd{7_r}{\kenv{(x \leftarrow \kmkd{7}{x})}{}{\kmkd{i_r}{E}}}}{ES}{FS}}}$ & \text{substituting in definition}\\
      \\
\end{longtable}
We'll use a symbol for this nested environment, as it is quite unwieldy.
\[
    E'' = \kmkd{6_r}{\kenv{(x \leftarrow \kmkd{\kval}{(\kmkd{\emptyset}{+}~1~2~\kmkd{\emptyset}{x})})}{\kmkd{7_r}{\kenv{(x \leftarrow \kmkd{7}{x})}{}{\kmkd{i_r}{E}}}}{\kmkd{i_r}{E}}}
\]
Let's continue along with the example.
\begin{longtable}{cc}
      \\
      $\kpval{\kmkd{\katt{\emptyset}{\emptyset}}{$&\text{evaluating body}\\
      $(\kprim{0}{eval}~\kmkd{\katt{\emptyset}{\emptyset}}{(\kprim{0}{array}~\kprim{1}{+}~\kmkd{6}{x}~\kmkd{6}{x})}~\kmkd{6}{de})}}{E''}{$
      &\\
      $ES}{FS}$ &\text{with generated env}\\
      \\
      $\kpval{\kmkd{\katt{\emptyset}{\emptyset}}{$&\text{evaluating array call}\\
      $(\kprim{0}{eval}~\kmkd{\kval}{(\kprim{1}{+}~(\kmkd{\emptyset}{+}~1~2~\kmkd{\emptyset}{x})(\kmkd{\emptyset}{+}~1~2~\kmkd{\emptyset}{x}))}$&\\
      $\kmkd{7_f}{\kenv{(x \leftarrow \kmkd{7}{x})}{}{\kmkd{i_r}{E}}})}}{E''}{$
      &\\
      $ES}{FS}$ & \text{and its symbol parameters}\\
      \\
      $\kpcombine{\kmkd{i''}{\kprim{-1}{veval}}}{$
      &\\
      $(\kunval{\kmkd{\kval}{(\kprim{1}{+}~(\kmkd{\emptyset}{+}~1~2~\kmkd{\emptyset}{x})(\kmkd{\emptyset}{+}~1~2~\kmkd{\emptyset}{x}))}}$
      &\\
      $\kmkd{7_f}{\kenv{(x \leftarrow \kmkd{7}{x})}{}{\kmkd{i_r}{E}}})}{$
      &\\
      $E''}{ES}{FS}$ & \text{evaluating eval}\\
 \\
      $\kpcombine{\kmkd{i''}{\kprim{-1}{veval}}}{$
      &\\
      $(\kmkd{\kfresh}{(\kprim{1}{+}~(\kmkd{\emptyset}{+}~1~2~\kmkd{\emptyset}{x})(\kmkd{\emptyset}{+}~1~2~\kmkd{\emptyset}{x}))}$
      &\\
      $\kmkd{7_f}{\kenv{(x \leftarrow \kmkd{7}{x})}{}{\kmkd{i_r}{E}}})}{$
      &\\
      $E''}{ES}{FS}$ & \text{unval}\\
      \\
      $\text{let}~\kmkd{7_f}{E'}~=~\kmkd{7_f}{\kenv{(x \leftarrow \kmkd{7}{x})}{}{\kmkd{i_r}{E}}}~\text{in}$ & \text{we also abbriviate E'}\\
      $\text{let}~V'~=~\kpval{$
      &\\
      $\kmkd{\kfresh}{(\kprim{1}{+}~(\kmkd{\emptyset}{+}~1~2~\kmkd{\emptyset}{x})(\kmkd{\emptyset}{+}~1~2~\kmkd{\emptyset}{x}))}}{$
      &\\
      $\kmkd{7_f}{E'}}{\{\kmkd{7_f}{E'}\}\cup ES}{FS}~\text{in}$ &\\
      $\kunder{V'}{(\kprim{-1}{veval}~V'~\kmkd{7_f}{E'})}{\kmkd{i_x}{E}}{ES}{FS}$ & \text{stepping veval}\\
\end{longtable}
V' will step to:
\[
    \text{let}~V'~=~\kmkd{\katt{\emptyset}{\emptyset}}{(\kprim{0}{+}~\kmkd{\katt{\emptyset}{\emptyset}}{(\kprim{0}{+}~3~\kmkd{7}{x})}~\kmkd{\katt{\emptyset}{\emptyset}}{(\kprim{0}{+}~3~\kmkd{7}{x})})}~\text{in}\\
\]
Now, the under call will fail to complete, as $returnOk$ will return false, finding $7$ in the result.
Thus, the fallback will be returned and will be the final result of the call to $double\_parameter$.
This time, $returnOk$ will be true, as $6$, $double\_parameter$'s ID, is nowhere to be found in the result.
Thus, we will transition to a call to $dropRV$ with the ID $7$.
\[
    dropRV(\kmkd{\katt{\emptyset}{\emptyset}}{(\kprim{-1}{veval}~V'~\kmkd{7_f}{E'})},~\kmkd{7_f}{E'},~ES,~ FS)
\]
Of course, this is a redundant veval, so the final result of partially evaluating the call to \\ $double\_parameter$ is:
\[
    V'
\]
that is,
\[
    \kmkd{\katt{\emptyset}{\emptyset}}{(\kprim{0}{+}~\kmkd{\katt{\emptyset}{\emptyset}}{(\kprim{0}{+}~3~\kmkd{7}{x})}~\kmkd{\katt{\emptyset}{\emptyset}}{(\kprim{0}{+}~3~\kmkd{7}{x})})}
\]

Note that this suspended computation is returned and replaces the call to the macro-like operative call.
This code is the equivalent of:
\begin{lstlisting}[language=Lisp,caption={},label=code:additionexample]
(vau (x) (+ (+ 3 x) (+ 3 x)))
\end{lstlisting}
This is what we would have gotten if we had used a macro and basic constant propagation.

Note that this evaluation of the example did not depend on the exact values, the code generated by the macro-like operative, or the parameters to the call.
In just this way, static calls to macro-like operatives will be partially evaluated away by our algorithm in a way congruent to macro-expansion and then constant propagation in a more standard optimizing Scheme implementation.

\section{Compiler Backend and Optimizations}
\label{sec:opt}

Partial evaluation can eliminate all inefficiencies from macro-like operatives but there are other inefficiencies left that require backend optimizations.
One major remaining inefficiency is dynamic combiner call sites.
In such cases, no local information is available ahead of time to determine if the combiner that will be called is an applicative or an operative - that is, whether the arguments will be evaluated or not and whether the combiner needs to access its calling dynamic environment.
This would normally mean that code in the parameter position of dynamic calls cannot even be partially evaluated.
To overcome this inefficiency, we tag the compiled combiner closure values with bits indicating their wrap level and need of the calling environment.
Each dynamic call site branches on these bits.
Inside the "wrap\_level=0" side of the dynamic branch, a reference to the unevaluated arguments written out in static memory is emitted.
Inside the "wrap\_level=1" side of the dynamic branch, the compiler re-invokes unval and the partial evaluator recursively on each argument, resulting in code that is again as efficient as a language without fexprs (plus the overhead of the dynamic branch).
Note that this requires that both the partial evaluation algorithm above as well as the compilation algorithm be extended to support failure. Failure during partial evaluation does not necessarily mean failure during run-time. For instance, a dynamic combiner might always be an operative with a wrap level of 0, and so some erroring parameter code is actually data, and will never be evaluated (and thus never lead to an error).

Garbage is collected via reference counting for simplicity. Since this is a purely functional language, there are no cycles to worry about.
In order to remove the rest of the inefficiencies after partial evaluation, we have implemented various compiler optimizations.
The following sections cover the other key optimizations implemented in our compiler. 

\subsection{Lazy Environment Instantiation}
We delay the allocation and initialization of dynamic environment values until they are actually needed.  Combiner calls that do take in the dynamic environment check a dedicated register to see if the environment value has already been created. If not, it creates it.
This means the dynamic execution traces of combiner calls where there is no call that takes in the dynamic environment never reifies it and incurs only a single (predictable) branch of overhead.
For static combiner calls, this information (if arguments are evaluated, if it takes in the surrounding environment, etc) is known at compile time, and no runtime branches are generated.

\subsection{Type-Inference-Based Primitive Inlining}
In order to reduce the overhead of every built-in operation being a combiner call with dynamic types, we implemented type-inference guided inlining of primitive operations.
An analysis pass infers types based on branch predicates, which works quite well with the code generated by our match operative.
For instance, the combiner \textit{len} can be inlined to just a few bit-twiddling opcodes by determining that a particular variable must contain an array in \textit{cond}.

For instance, consider the following code:
    \begin{lstlisting}[language=Lisp,caption={Type Inference Example},label=code:typeinfer]
(cond (and (array? a) (= 3 (len a)))    (idx a 2)
      true                              nil)\end{lstlisting}
The call to \textit{idx} can be fully inlined without type or bounds checking because it resides in a block only reachable if the variable 'a' does contain an array of length 3.
No type information is needed to inline type predicates, as they only need to look at the tag bits.
Equality checks can be inlined as a simple word/ptr compare if any of its parameters are of a type that can be word/ptr compared (ints, bools, and symbols).
When type inference and primitive inlining is combined together, it means that every primitive call in most match expressions can be fully inlined into a handful of opcodes apiece.
In the above example, every single primitive listed will be inlined: the \textit{cond} to WebAssembly if blocks, the predicate functions to bit-twiddling and branches, the \text{idx} to bit-twiddling and a load with a constant offset, etc.

\subsection{Immediately-Called Closure Inlining}
Inlining calls to closure values that are allocated and then immediately used helps incur no overhead for implementing some operatives. The main macro-like operative reaping the benefit is "let".
As seen below, Listing \ref{code:letinline} is partially evaluated to the equivalent of Listing \ref{code:letinline2}, then inlined. As a result, the only overhead is the creation of a new environment, which is further made lazy and eliminated in the common case by Lazy Environment Instantiation. In this way, "let" is actually syntactic sugar for the definition and immediate call of a closure, like in many lambda calculi, but no efficiency is lost by doing so.
\begin{lstlisting}[language=Lisp,caption={Let Inlining Example},label=code:letinline]
    (let (a (+ 1 2))
         (+ a 3))
\end{lstlisting}
\begin{lstlisting}[language=Lisp,caption={Let Inlining Example - Expanded},label=code:letinline2]
    ((wrap (vau (a) (+ a 3))) (+ 1 2))
\end{lstlisting}

\subsection{Y-Combinator Elimination}
Continuing the theme of making the classic lambda-calculi implementations of concepts as efficient as standard implementations, the final set of optimizations ensures no overhead from using the Y-Combinator to implement recursion.
In Kraken, the Y-Combinator looks like  Listing \ref{code:ycomb} with a tiny example of its use shown in Listing \ref{code:ycombuse}.
    \begin{lstlisting} [language=Lisp,caption={The Y Combinator, as defined in Kraken},label=code:ycomb]
(let Y (lambda (f)
           ((lambda (x) (x x))
            (lambda (x) (f (wrap (vau app_env (& y) (lapply (x x) y app_env)))))))
)\end{lstlisting}

    \begin{lstlisting}[language=Lisp,caption={A Factorial function explicitly using the Y Combinator},label=code:ycombuse]
(Y (lambda (recurse) (lambda (n) (if (= 0 n) 1
                                             (* n (recurse (- n 1)))))))\end{lstlisting}

Normally, one does not manually use the Y-Combinator. In Kraken, there is a \textit{rec-lambda} derived operative that is easy to use and evaluates Y-Combinator behind the scenes.
Y-Combinator is actually always used to implement recursion in Kraken whether explicitly stated or not.
This allows us to keep our language pure, and in agreement with the calculus.

This optimization actually falls out naturally from our architecture, with just a little bit of care taken while bookkeeping.
When compiling a combiner, the compiler first inserts what the combiner index will be into a memoization dictionary before re-executing partial evaluation on the body of the combiner.
Any static recursive calls will have the exact form of the combiner currently being compiled, and so the compiler can emit a static reference to the correct combiner index.
All of this works because the re-executed partial evaluation of the body before compilation made sure to normalize the form of the combiner of a recursive call to be identical to that of the combiner being compiled before this partial-evaluation.
Since this is an eager language, the definition of the Y-Combinator in our language has an extra closure to prevent infinite recursion inside the Y-Combinator itself.
We, thus, additionally implement eta-conversion in the compiler to remove this extra level of indirection. Since the expression inside is now a constant instead of a call, there is no risk of infinite recursion. Combined with the normalization above, we achieve fully-efficient static recursive calls when using the Y-Combinator to define recursive functions.

Finally, as a purely functional Lisp, we use recursion instead of iteration.
While we wait for the tail\_call instruction in WebAssembly to be merged and implemented, we implemented a more limited form of Tail Call Elimination where auto-recursive calls in tail position are transformed into branches to the head of a loop that encloses the combiner's body.
The combination of Tail Call Elimination with Y-Combinator Elimination above means that a recursive function defined using the Y-Combinator can be as efficient as an imperative loop in other languages.
When WebAssembly finishes implementing the tail\_call instruction, it can easily be emitted to gain full proper tail calls.

 \section{Benchmarks and Evaluation}
\label{sec:eval}

We evaluate \krakenSpace to answer the following set of questions:
\begin{itemize}
\item How much improvement does partial evaluation and our implemented compiler optimizations give \kraken? 
\item How much faster is our purely functional f-expr language, \krakenSpace, compared to other implementations of fexprs? 
\item How does \kraken's performance, with its fexprs, compare to macros? 
\item How do the different partial evaluation mechanisms/optimizations in \krakenSpace contribute towards reduction in overall runtime?
\end{itemize}

\textbf{Experimental Setup}: 
We ran these experiments in a reproducible Nix environment on a NixOS install \cite{10.1145/1411203.1411255} (Kernel 6.0.0) on a laptop with 8 cores / 16 threads and 64 GB of RAM.
Our code contains the scripts and Nix Flakes needed to reproduce the exact set of dependencies to run our tests.

The Kraken benchmarks were run using both the Wasmtime and WAVM WebAssembly engines for most benchmarks.
The Wasmtime WebAssembly engine is one of the most popular, developed by the Bytecode Alliance itself, and uses the CraneLift code generation backend.
The WAVM WebAssembly engine is interesting for its use of LLVM, and it often produces the fastest code on benchmarks but has a higher startup time.
We eliminated the Cfold Wasmtime benchmark due to problems running out of stack space (a known property of the Cfold benchmark).

\textbf{Benchmarks}: 
To showcase the capability of Kraken, we created benchmarks that are commonly implemented in functional languages and have been used as benchmarks in other papers \cite{reinking2021perceus, 10.1145/3547646}.
The benchmarks are
\begin{itemize}
\item Fib - Calculating the nth Fibonacci number
\item RB-Tree - Inserting n items into a red-black tree, then traversing the tree to sum its values
\item Deriv - Computing a symbolic derivative of a large expression
\item Cfold - Constant-folding a large expression
\item NQueens - Placing n number of queens on the board such that no two queens are diagonal, vertical, or horizontal from each other
\end{itemize}
All benchmarks besides Fibonacci use the fexpr version of match for pattern matching in \kraken, which is equivalent to the macro version in NewLisp. We also RB-Tree using NewLisp's~\cite{mueller2018newlisp} version of fexpr match. We modified the sizes of the problems presented to the benchmark to account for the longer running times of some of the less-optimized implementations.
The code for Kraken and NewLisp is very similar, and we should note that it is very unidiomatic NewLisp.
Our goal was not to compare Kraken and NewLisp as implementation languages for Red-Black Trees, but to stress test a single reasonably complex fexpr/macro, namely pattern matching.

\subsection{The Effect of Partial Evaluation on Eval Calls}

\begin{table}[h]
\caption{Number of eval calls with no partial evaluation for Fexprs}
	\begin{tabular}{||c | c c c c c ||} 
		\hline
		&Evals & Eval w1 Calls & Eval w0 Calls & Comp Dyn & Comp Dyn\\ 
        & & & & w1 Calls & w0 Calls\\ [0.5ex] 
		\hline\hline
		Cfold 5 & 10897376 & 2784275 & 879066  & 1 & 0 \\ 
		\hline
		  Deriv 2  & 11708558 & 2990090 & 946500 & 1 & 0 \\ 
        \hline
		  NQueens 7 & 13530241 & 3429161 & 1108393 & 1 & 0 \\ 
    \hline
		  Fib 30 & 119107888 & 30450112 & 10770217 & 1 & 0 \\ 
    \hline
		  RB-Tree 10 & 5032297 & 1291489 & 398104 & 1 & 0 \\ 
		\hline
	\end{tabular}
    \label{npe:calls}
 \end{table}

As mentioned before, using fexprs without partial evaluation will prelude optimization and cause a massive amount of repeated work. Table \ref{npe:calls} and Table \ref{pe:calls} show the number of calls to the \krakenSpace runtime's eval function, the number of times the runtime's eval function executed a call to an applicative with wrap\_level=1, the number of times the runtime's eval function executed a call to an operative with wrap\_level=0, the number of compiled dynamic calls to applicatives with wrap\_level=1, and the number of compiled dynamic calls to operatives with wrap\_level=0.
These are shown for \krakenSpace test cases with partial evaluation turned off and turned on. 
\begin{table}[h]
\caption{Number of eval calls in Partially Evaluated Fexprs}
	\begin{tabular}{||c | c c c c c ||} 
		\hline
		&Evals & Eval w1 Calls & Eval w0 Calls & Comp Dyn & Comp Dyn\\ 
        & & & & w1 Calls & w0 Calls\\ [0.5ex] 
		\hline\hline
		Cfold 5 & 0 & 0 & 0  & 0 & 0 \\ 
		\hline
		  Deriv 2  & 0 & 0 & 0 & 2 & 0 \\ 
        \hline
		  NQueens 7 & 0 & 0 & 0 & 0 & 0 \\ 
    \hline
		  Fib 30 & 0 & 0 & 0 & 0 & 0 \\ 
    \hline
		  RB-Tree 10 & 0 & 0 & 0 & 10 & 0 \\ 
		\hline
	\end{tabular}
    \label{pe:calls}
 \end{table}

\begin{table}[h]
\caption{Number of calls to the runtime's eval function for RB-Tree. The table shows the non-partial evaluation numbers -> partial evaluation numbers.}
	\begin{tabular}{||c | c c c c c ||} 
		\hline
		&Evals & Eval w1 Calls & Eval w0 Calls & Comp Dyn & Comp Dyn\\ 
        & & & & w1 Calls & w0 Calls\\ [0.5ex] 
		\hline\hline
		  RB-Tree 7 & 2952848 -> 0 & 757932 -> 0 & 233513 -> 0 & 1 -> 7 & 0 -> 0\\ 
        \hline
		  RB-Tree 8 & 3532131 -> 0 & 906548 -> 0 & 279379 -> 0 & 1 -> 8 & 0 -> 0\\ 
        \hline
		  RB-Tree 9 & 4278001 -> 0 & 1097965 -> 0 & 3383831 -> 0 & 1 -> 9 & 0 -> 0\\ 
		\hline
	\end{tabular}
    \label{pe:rb}
    \vspace{-4mm}
 \end{table}

Without partial evaluation, no compilation can be done because it is impossible to tell if arguments to calls will be evaluated. In all benchmarks, partial evaluation removed all calls to the runtime's eval function, resulting in a completely compiled program. Looking at RB-Tree, there are over a million calls to combiners with wrap level 1 (normal functions), and 398,000 calls to combiners with wrap level 0 (operatives replacing macros). This massive blowup in the number of calls is due to the repeated and exponential re-execution of macro-like-combiners in the definition of other macro-like-combiners, as discussed in the Introduction.

The non-partially-evaluated benchmarks show 1 compiled dynamic call to an applicative (its the first call into eval) and 0 compiled dynamic calls to operatives, because there is no compilation at all. For the partially evaluated benchmarks, there are a few compiled dynamic calls to applicatives due to higher-order function use in the benchmarks, and there are no compiled dynamic calls to operatives, as all operative use has been eliminated.
We also varied the inputs for RB-Tree shown in Table \ref{pe:rb} to give a sense for how the number scale with respect to input size.

The incredible slowdown implied by these tables comes to full fruition in our RB-Tree test in Fig.~\ref{fig:kraken_nqueens_rbtree}.
We kept this run shorter because Kraken's non-partial-evaluating interpreter takes an incredibly long time even for 100 insertions (40 minutes).
The compounding layers of repeated macro-like operative calls in the non-partially-evaluated Kraken version cause a ~70,000x slowdown relative to the partial evaluated, optimized, and compiled version.
For the remaining benchmarks, we remove the naive interpreted \krakenSpace version, as in each case its performance is so bad as to blow out the graph and make it impossible to do any comparison.
In our optimized Kraken, our partial evaluation algorithm is able to fully collapse these levels of inefficiency, evaluate and inline the results, and give the backend more specialized code to optimize, emitting a compiled version that handily beats not only the NewLisp-fexpr implementation but even the NewLisp-macro implementation, as can be seen in Fig.~\ref{fig:kraken_vs_world_fib}.
We kept the benchmark sizes small in this test because the stack limits of NewLisp prevent sizes larger then ~880, while the Tail Call Elimination performed by the \krakenSpace compiler allows us to run much larger benchmarks, including the run of 4,800,000 inserts to the RB-Tree.
This result shows the dramatic effect of partial evaluation and compiler optimizations on runtime for \kraken. Our technique takes the performance of a fully fexpr based language from being completely infeasible to being faster than a macro-based dynamic scripting language currently in use.

\begin{figure}[h]
\caption{Constant Fold and Deriv}
\includegraphics[width=0.45\textwidth]{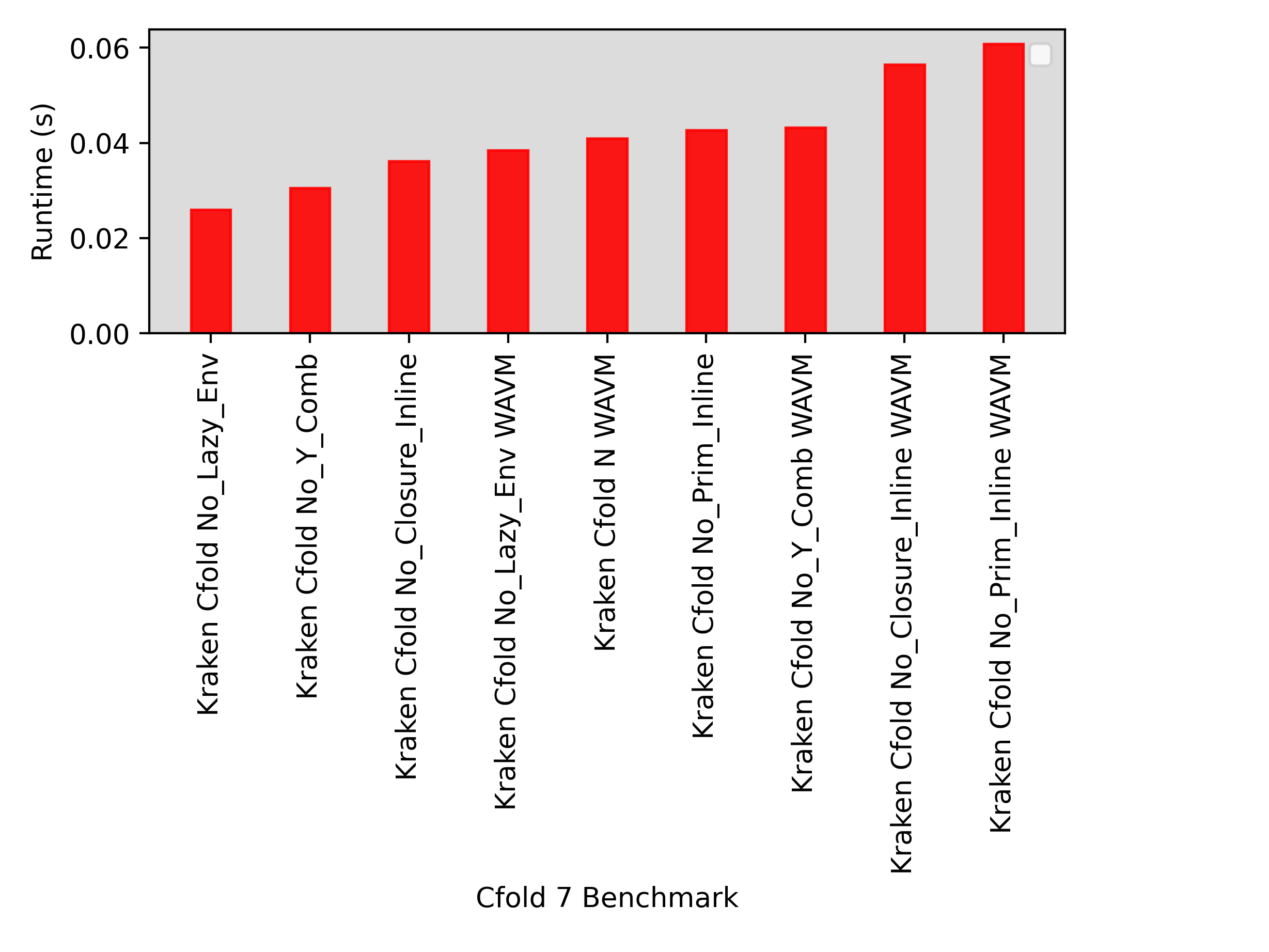}
\includegraphics[width=0.45\textwidth]{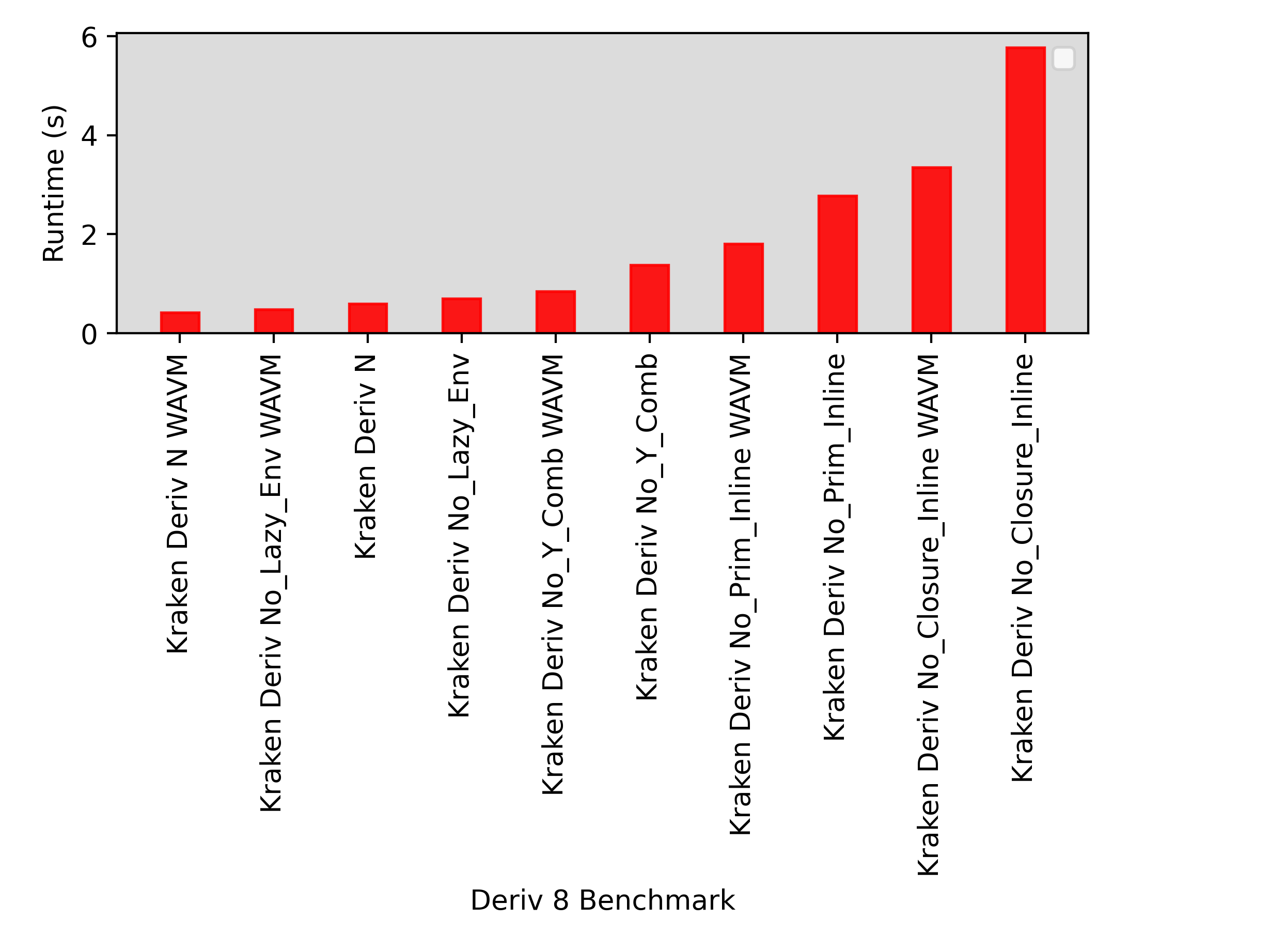}
\label{fig:kraken_const_deriv}
\vspace{-6mm}
\end{figure}
\subsection{Comparison between Kraken Versions}
\begin{figure}[h]
\caption{N-Queens}
\includegraphics[width=0.45\textwidth]{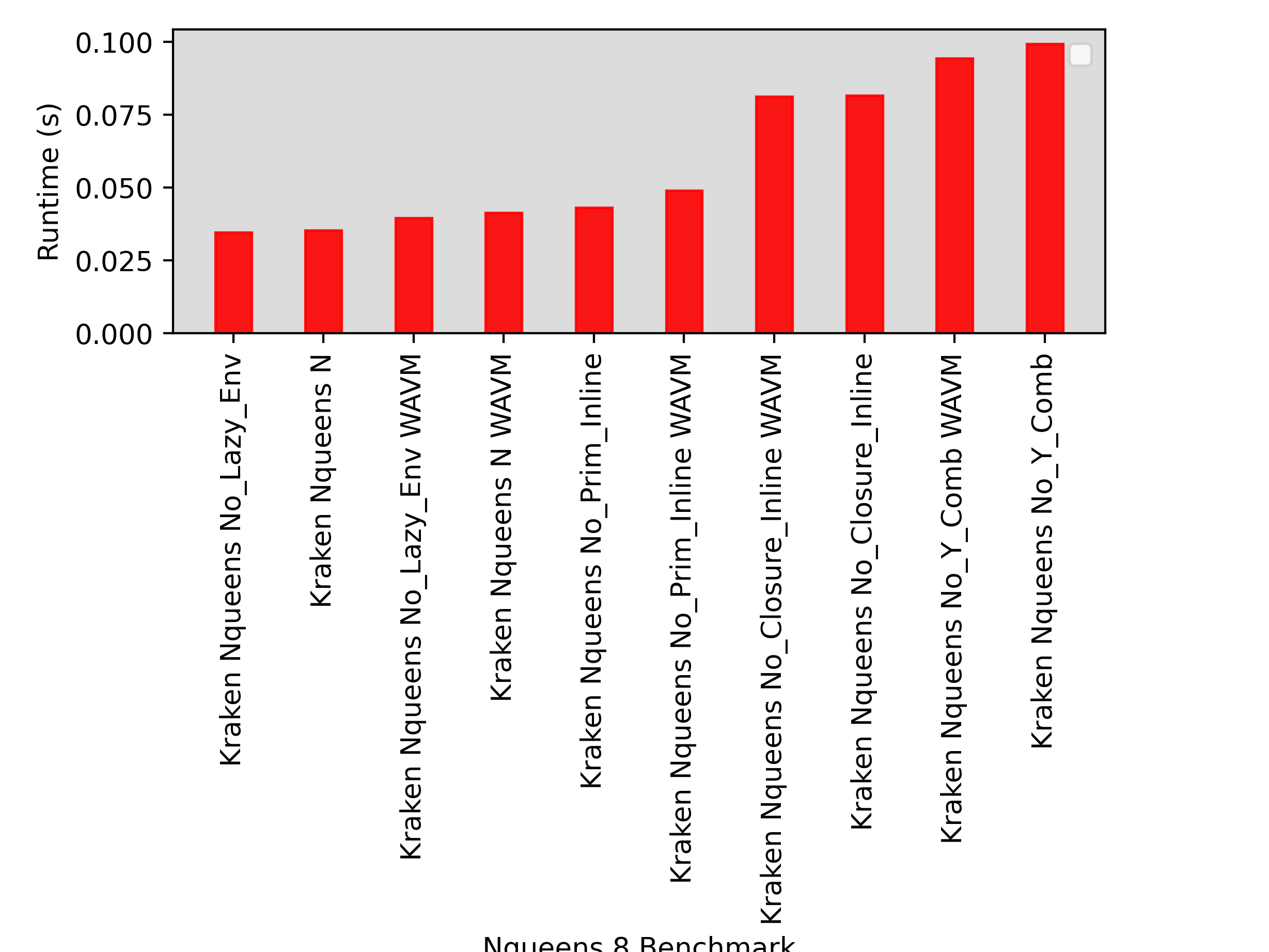}
\includegraphics[width=0.45\textwidth]{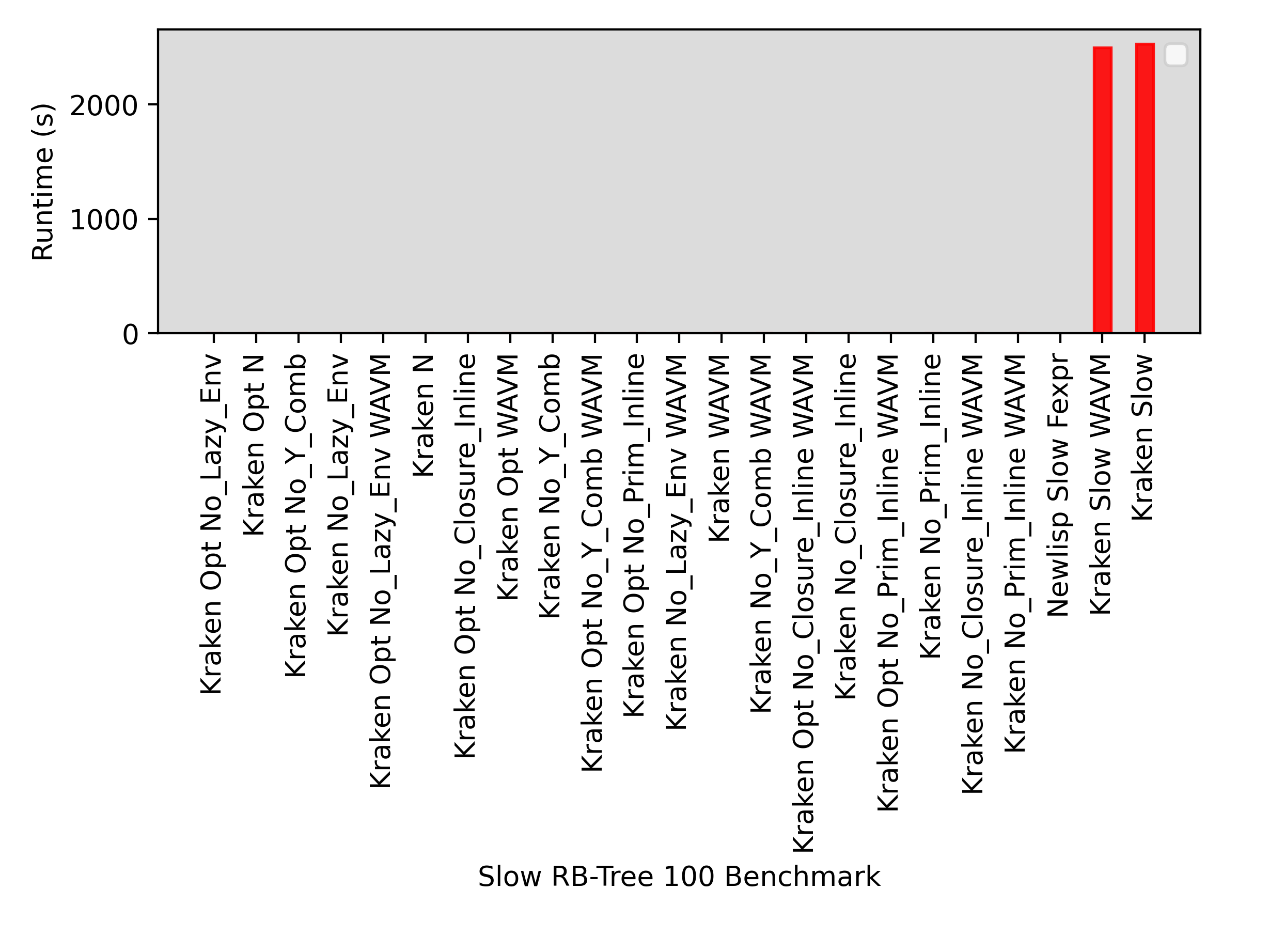}
\label{fig:kraken_nqueens_rbtree}
\vspace{-4mm}
\end{figure}

Beyond the massive speedup from partial-evaluation, Fig. \ref{fig:kraken_const_deriv} and \ref{fig:kraken_nqueens_rbtree} show the effect of the various compiler optimizations we described by disabling them one by one.
 Our main four optimizations have a strong positive effect on runtime, with the exception of lazy environment instantiation. Lazy environment instantiation helps massively on fib, and some on Deriv, but generally hurts the rest slightly.

\subsection{Comparison against Others}

\begin{figure}[h]
\caption{Kraken vs. Others. Ordered by fastest to slowest}
\includegraphics[width=0.45\textwidth]{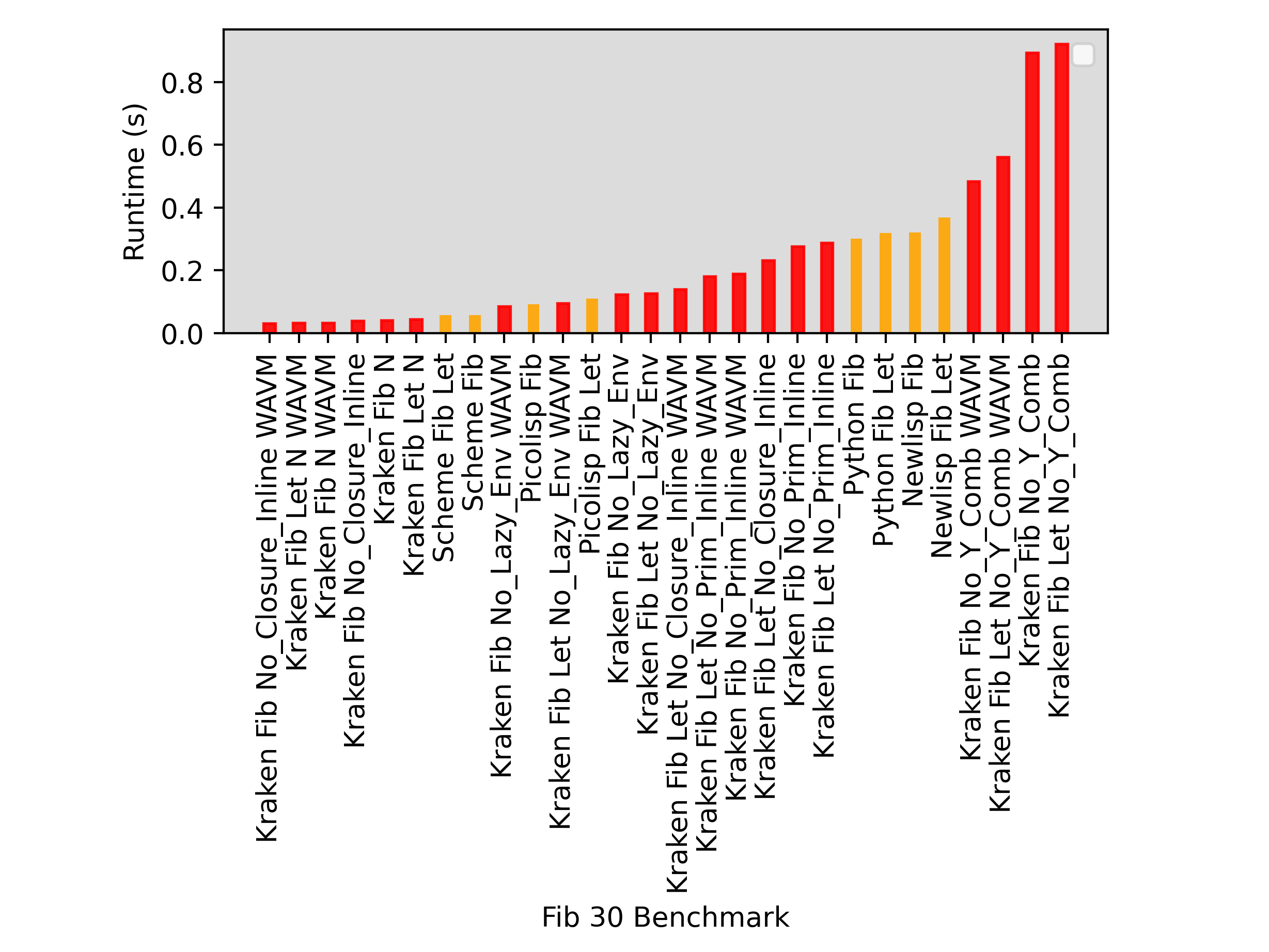}
\includegraphics[width=0.45\textwidth]{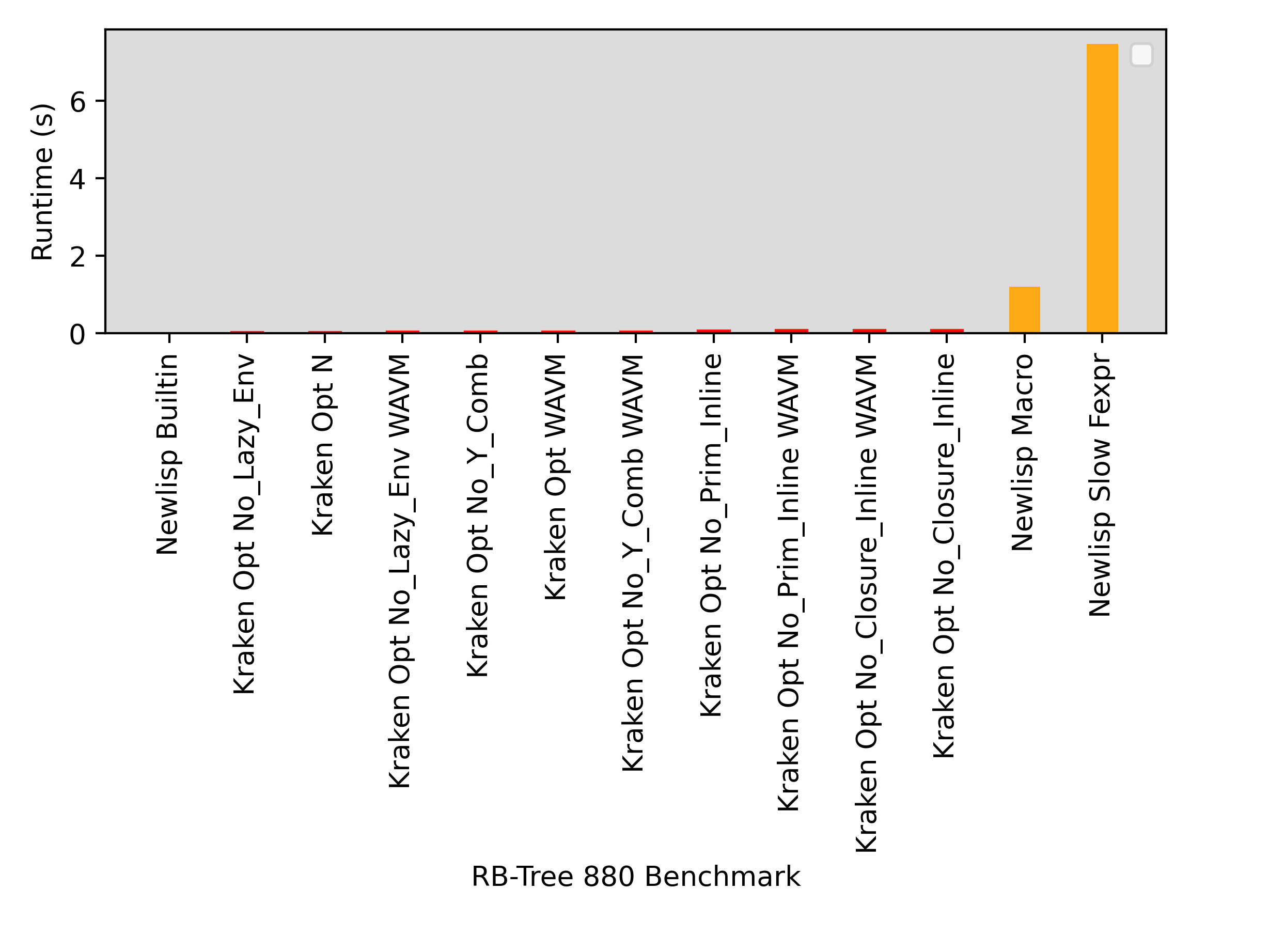}
\label{fig:kraken_vs_world_fib}
\end{figure}

To give a general idea of our current performance, we also show a Fibonacci benchmark that mostly exercises pure function-call speed and inlining as seen in Fig. ~\ref{fig:kraken_vs_world_fib}.
We include Python and Chez Scheme to give a general idea for where an exemplar slow and an exemplar fast dynamic language would fall.
With the benefit of our partial evaluation, compilation, and leaning upon mature WebAssembly implementations, we beat both, but this should be taken with a grain of salt, as this is a very limited micro-benchmark only meant to give a general sense of the order of magnitude of our performance.

\label{sec:eval1}

\section{Related Work}
\label{sec:relatedwork}

The focus of this paper has been on efficient fexpr language implementation using partial evaluation.

\textbf{Other Fexpr Implementations:}

Aside from NewLisp~\cite{mueller2018newlisp}, picoLisp~\cite{burger2013picolisp} also includes fexprs as a language feature.
Similar to NewLisp, picoLisp is a dynamic and interpreted language and does no special optimization of fexprs. Like NewLisp, most forms in picoLisp are implemented in the interpreter itself instead of being built up from smaller components, as is common to do with macros in other Lisps that use macros.
While both of these languages originally lacked a macro implementation, NewLisp eventually added one due to the confusing evaluation rules and performance problems caused by its fexprs.

However, John Shutt demonstrated in his 2010 thesis \cite{shutt2010fexprs} that if fexprs are re-formulated divorced from their historical context and co-existence with other now-extinct language features, like dynamic scoping, they can be an elegant and well-behaved alternative to macros. It is important, however, that the language be designed with this in mind from the start. We have additionally shown that the performance problems that have plagued fexprs are solvable, fixing the other main issue that has kept fexprs from common usage.

\textbf{Partial Evaluation:}

Our work's main contribution is the partial evaluation scheme that provides the basis for the performance improvement of our fexpr implementation.
Partial evaluation has been around for many years, including during the early years of Lisp \cite{lombardi1964lisp}, and is utilized in many languages.
Jones~\cite{10.1145/243439.243447} and Charles \& Olivier~\cite{10.1145/158511.158707} provide a high-level look at partial evaluation and its history in the community and explain well the differences between online and offline techniques.
Ruf provides details of interesting online partial evaluation techniques in his thesis~\cite{ruf1993topics}, and Sperber \& Thiemann~\cite{10.1145/249069.231419}~give a description of the technique of using static data descriptions as a stand in for dynamic data in their paper on compilation using partial evaluation, which is similar to our handling of partially-static-data environments.
Danvy, Malmkj{\ae}r, and Palsberg use eta-expansion as a part of partial evaluation in their implementation of "The Trick" in \cite{danvy1996eta}, whereas we use it during compilation to optimize the remnents of the Y-Combiner.
Partial evaluation's application is broad, having been utilized for many imperative languages like C~\cite{andersen1992self}, Matlab~\cite{elphick2003partial} and Pascal~\cite{meyer1991techniques}, as well as for other programming paradigms, such as logic programming, where Lloyd \& Sheperdson~\cite{LLOYD1991217} provided a theoretical foundation that got expanded by Alpuente et al~\cite{alpuente1998partial}.
Romph et al. use partial evaluation/staging to optimize data structures in Scala in \cite{rompf2013optimizing}, Sperber \& Thiemann further use partial evaluation in the generation of LR parsers in \cite{sperber2000generation}, and Sestoft uses partial evaluation to compile efficient ML pattern matching \cite{sestoft1996ml}.



\section{Conclusion and Future Work}
\label{sec:conclusion}
In this work, we proposed a purely functional Lisp based on fexprs, \kraken, and the first-ever compilation framework to make fexprs performant by utilizing partial evaluation.
The partial evaluator will evaluate away all fexprs that behave like macros (namely operatives of a particular form), showing that our marcro-esque fexprs can be as efficient as compile-time macros with a minor penalty to dynamic function calls.
In addition, the partial evaluator with the compiler optimizations produced better-optimized code than the base interpreter (70,000x faster) or NewLisp's fexpr implementation~\cite{mueller2018newlisp} (233x faster).
We have shown that fexprs can be a reasonably performant alternative to macros and are a viable foundation for new expressive functional languages in the Lisp tradition.

In the future we will look at improving the partial evaluator to handle more flexible definitions of operatives that are less strictly macro-like, as well as investigation into adding features such has delimited continuations, which we had removed from Shutt's Vau calculi for simplicity.



\bibliography{bibfile}

\appendix

\section{Auxiliary Helper Relations}
\begin{figure}[H]
\[
  \begin{array}{rcll}
    \kprog{n}      &=& \emptyset\\
    \kprog{\kmkd{\emptyset}{s}}  &=& \emptyset\\
    \kprog{\kmkd{i}{s}} &=& \{i\}\\
    \kprog{\kmkd{\kval}{(T_1~\dots~T_n)}} &=& \kprog{T_1} \cup \dots \cup \kprog{T_n}\\
    \kprog{\kmkd{\kfresh}{(T_1~\dots~T_n)}} &=& \{\truev\}\\
    \kprog{\kmkd{\katt{x}{y}}{(T_1~\dots~T_n)}} &=& \{x\} \cup \kprog{T_1} \cup \dots \cup \kprog{T_n}\\
    \kprog{\kmkd{x}{\kcomb{n}{s'}{\kmkd{i}{E}}{(s\dots)}{T}}} &=& (\kprog{\kmkd{i}{E}} \cup \kprog{T}) - \{x\}\\
    \kprog{\kprim{n}{o}} &=& \emptyset\\
    \kprog{\kmkd{x}{\kenv{(s \leftarrow T)\dots}{s' \leftarrow \kmkd{i'}{E'}}{\kmkd{i}{E}}}} &=& \kprog{\kmkd{i}{E}} \cup \kprog{\kmkd{i'}{E'}}\\
    \kprog{\kmkd{x}{\kenv{(s \leftarrow T)\dots}{}{\kmkd{i}{E}}}} &=& \kprog{\kmkd{i}{E}}\\
  \end{array}
  \]
  \caption{Needed-For-Progress Relation}
  \label{fig:neededforprogressrelation}
\end{figure}
\begin{figure}[H]
\[
  \begin{array}{rcll}
    \kuprog{n}      &=& \emptyset\\
    \kuprog{\kmkd{x}{s}} &=& \emptyset\\
    \kuprog{\kmkd{\kval}{(T_1~\dots~T_n)}} &=& \kuprog{T_1} \cup \dots \cup \kuprog{T_n}\\
    \kuprog{\kmkd{\kfresh}{(T_1~\dots~T_n)}} &=& \kuprog{T_1} \cup \dots \cup \kuprog{T_n}\\
    \kuprog{\kmkd{\katt{x}{y}}{(T_1~\dots~T_n)}} &=& \kuprog{T_1} \cup \dots \cup \kuprog{T_n}\\
    \kuprog{\kmkd{x}{\kcomb{n}{s'}{\kmkd{i}{E}}{(s\dots)}{T}}} &=& \emptyset\\
    \kuprog{\kprim{n}{o}} &=& \emptyset\\
    \kuprog{\kmkd{x}{\kenv{(s \leftarrow T)\dots}{s' \leftarrow \kmkd{i'}{E'}}{\kmkd{i}{E}}}} &=& \kuprog{\kmkd{i}{E}} \cup \kuprog{\kmkd{i'}{E'}}\\
    && \text{if}~\kprog{\kmkd{x}{\kenv{(s \leftarrow T)\dots}{\kmkd{i'}{E'}}{\kmkd{i}{E}}}} = \emptyset\\
    \kuprog{\kmkd{x}{\kenv{(s \leftarrow T)\dots}{}{\kmkd{i}{E}}}} &=& \kuprog{\kmkd{i}{E}}\\
    && \text{if}~\kprog{\kmkd{x}{\kenv{(s \leftarrow T)\dots}{}{\kmkd{i}{E}}}} = \emptyset\\

    \kuprog{\kmkd{x}{\kenv{(s \leftarrow T)\dots}{s' \leftarrow \kmkd{i'}{E'}}{\kmkd{i}{E}}}} &=& \{x\} \cup \kuprog{\kmkd{i}{E}} \cup \kuprog{\kmkd{i'}{E'}}\\
    && \text{if}~\kprog{\kmkd{x}{\kenv{(s \leftarrow T)\dots}{\kmkd{i'}{E'}}{\kmkd{i}{E}}}} \neq \emptyset\\
    \kuprog{\kmkd{x}{\kenv{(s \leftarrow T)\dots}{}{\kmkd{i}{E}}}} &=& \{x\} \cup \kuprog{\kmkd{i}{E}}\\
    && \text{if}~\kprog{\kmkd{x}{\kenv{(s \leftarrow T)\dots}{}{\kmkd{i}{E}}}} \neq \emptyset\\
  \end{array}
  \]
  \caption{Needed-For-Progress-Upper Relation}
  \label{fig:neededforprogressupperrelation}
\end{figure}

\begin{figure}[H]
\[
  \begin{array}{rcll}
    \kiprog{n}      &=& \emptyset\\
    \kiprog{\kmkd{x}{s}} &=& \emptyset\\
    \kiprog{\kmkd{\kval}{(T_1~\dots~T_n)}} &=& \kiprog{T_1} \cup \dots \cup \kiprog{T_n}\\
    \kiprog{\kmkd{\kfresh}{(T_1~\dots~T_n)}} &=& \kiprog{T_1} \cup \dots \cup \kiprog{T_n}\\
    \kiprog{\kmkd{\katt{x}{y}}{(T_1~\dots~T_n)}} &=& \{y\} \cup \kiprog{T_1} \cup \dots \\
    && \cup~\kiprog{T_n}\\
    \kiprog{\kmkd{x}{\kcomb{n}{s'}{\kmkd{i}{E}}{(s\dots)}{T}}} &=& \emptyset\\
    \kiprog{\kprim{n}{o}} &=& \emptyset\\
    \kiprog{\kmkd{x}{\kenv{(s \leftarrow T)\dots}{s' \leftarrow \kmkd{i'}{E'}}{\kmkd{i}{E}}}} &=& \emptyset\\
    \kiprog{\kmkd{x}{\kenv{(s \leftarrow T)\dots}{}{\kmkd{i}{E}}}} &=& \emptyset\\
  \end{array}
  \]
  \caption{Needed-For-Progress-Resume-Infinite Relation}
  \label{fig:neededforprogressresumeinfiniterelation}
\end{figure}

\begin{figure}[H]
\[
  \begin{array}{rcll}
    lookup(s,&&\\\kenv{(s_1 \leftarrow V_1)(s_2 \leftarrow V_2)\dots}{s' \leftarrow \kmkd{i_x}{E}}{\kmkd{i'_{x'}}{E'}}) &=& lookup(s,\kenv{(s_2 \leftarrow V_2)\dots}{s' \leftarrow \kmkd{i_x}{E}}{\kmkd{i'_{x'}}{E'}})\\
    lookup(s,\kenv{(s \leftarrow V)\dots}{s' \leftarrow \kmkd{i_x}{E}}{\kmkd{i'_{x'}}{E'}}) &=& V\\
    lookup(s,\kenv{}{s \leftarrow \kmkd{i_x}{E}}{\kmkd{i'_{x'}}{E'}}) &=& \kmkd{i_x}{E}\\
    lookup(s,\kenv{}{s' \leftarrow \kmkd{i_x}{E}}{\kmkd{i'_{x'}}{E'}}) &=& lookup(s,E')\\
  \end{array}
  \]
  \caption{lookup}
  \label{fig:lookup}
\end{figure}

\begin{figure}[H]
\[
  \begin{array}{rcll}
    returnOk(n,i) &=& \truev\\
    returnOk(o,i) &=& \truev\\
    returnOk(\kmkd{\emptyset}{s},i) &=& \truev\\
    returnOk(\kmkd{\kval}{(V\dots)},i) &=& \truev\\
    returnOk(\kmkd{i'}{C},i) &=& \lnot IDin(i,\kmkd{i'}{C})\\
    returnOk(E,i) &=& \lnot IDin(i,E)\\
    \\

    returnOk(\kmkd{y}{(\kprim{-1}{veval}~z~\kmkd{i'_{x'}}{E})},i) &=& returnOk(\kmkd{i'_{x'}}{E},i)\\

        %
    returnOk(\kmkd{y}{(f~V\dots)},i) &=& \lnot takesDE(f) \land \bigwedge returnOk(V,i)\dots\\
  \end{array}
  \]
  \caption{returnOk}
  \label{fig:returnok}
\end{figure}
IDin (fg. \ref{fig:idin}) is quite simple, but pulled out to make returnOk more readable:
\begin{figure}[H]
\[
  \begin{array}{rcll}
    IDin(x,i) &=& i \in \kprog{x} \lor i \in \kuprog{x}\\
  \end{array}
  \]
  \caption{IDin}
  \label{fig:idin}
\end{figure}
"takesDE" (fg. \ref{fig:takesDE}) is similarly simple, returning if the primitive or derived combiner takes in its dynamic environment.
\begin{figure}[H]
\[
  \begin{array}{rcll}
    takesDE(\kprim{0}{vau}) &=& \truev\\
    takesDE(\kprim{0}{if0}) &=& \truev\\
    takesDE(\kprim{-1}{vif0}) &=& \truev\\
      takesDE(\kprim{x}{y}) &=& \falsev & \text{(Other primitives don't)}\\
    takesDE(\kmkd{i''}{\kcomb{n}{s'}{\kmkd{i'_r}{E'}}{(s\dots)}{Tb}}) &=& s' \neq \emptyset\\
      takesDE(x) &=& \truev \quad& \text{(any other term, including suspended terms,}\\
               &&&\text{we count as true to be safe)}\\
  \end{array}
  \]
  \caption{takesDE}
  \label{fig:takesDE}
\end{figure}

dropRV handles two main cases - the first is a call to veval, which is removed if it is redundant (if the ID of the explicit environment matches the ID of the current dynamic environment).
The second is a suspended function call, which calls dropRV recursively on all parameters.
If this does change the call, then we re-partially evaluate it, as perhaps with simpler parameters it can now be evaluated further.

\begin{figure}[H]
\[
  \begin{array}{rcll}
    dropRV(\kmkd{y}{(\kprim{-1}{veval}~z~\kmkd{i_f}{E})},\kmkd{i_x}{E},ES,FS) &=& dropRV(z,\kmkd{i_x}{E},ES,FS)\\
    &&(\text{only if}~y \neq \kval)\\
    \\

    dropRV(\kmkd{y}{(\kprim{n}{o}~V\dots)}, \kmkd{i_x}{E},ES,FS) &=&\textbf{let}~g~=~(\kprim{n}{o}~V\dots)\\
                                                               &&\textbf{let}~z~=~(\kprim{n}{o}~dropRV(V,\kmkd{i_x}{E},ES,FS)\dots)~\textbf{in}\\
                                                               &&\kpval{\kmkd{\kfresh}{z},\kmkd{i_x}{E}}{\kmkd{i_x}{E}}{ES}{FS}~\textbf{if}~z~\neq g\\
                                                               &&\textbf{else}~\kmkd{y}{z}\\
    &&(\text{only if}~y \neq \kval, n \neq -1)\\

    dropRV(&&\\
         \kmkd{y}{(\kmkd{i''}{\kcomb{n}{s'}{\kmkd{i'_r}{E'}}{(s\dots)}{Tb}}~V\dots)},&&\\\kmkd{i_x}{E},ES,FS) &=& \textbf{let}~g~=~(\kmkd{i''}{\kcomb{n}{s'}{\kmkd{i'_r}{E'}}{(s\dots)}{Tb}}~V\dots)~\textbf{in}\\
                                                               &&\textbf{let}~z~=~(\kmkd{i''}{\kcomb{n}{s'}{\kmkd{i'_r}{E'}}{(s\dots)}{Tb}}\\&&\qquad\qquad\qquad dropRV(V,\kmkd{i_x}{E},ES,FS)\dots)~\textbf{in}\\
                                                               &&\kpval{\kmkd{\kfresh}{z},\kmkd{i_x}{E}}{\kmkd{i_x}{E}}{ES}{FS}~\textbf{if}~z~\neq g\\
                                                               &&\textbf{else}~\kmkd{y}{z}\\

    \\
    dropRV(z,\kmkd{i_x}{E},ES,FS) &=& z\quad\text{Otherwise, return unchanged}\\
  \end{array}
  \]
  \caption{dropRV}
  \label{fig:droprv}
\end{figure}

\begin{figure}[H]
\[
  \begin{array}{rcl}
    \kpcombine{\kprim{0}{eval}}{(V_1~V_2)}{\kmkd{i_x}{E}}{ES}{FS} &\rightarrow& \kpcombine{\kmkd{i''}{\kprim{-1}{veval}}}{\\&&\qquad(\kunval{V_1}~V_2)}{\\&&\kmkd{i_x}{E}}{ES}{FS}\\
    \kpcombine{\kprim{-1}{veval}}{(V~\kmkd{i'_{x'}}{E'})}{\kmkd{i_x}{E}}{ES}{FS} &\rightarrow&\textbf{let}~V'~=\kpval{V}{\kmkd{i'_{x'}}{E'}}{(\kmkd{i'_{x'}}{E'} \cup ES)}{FS}~\textbf{in}\\&&\kunder{V'}{(\kprim{-1}{veval}~V'~\kmkd{i'_{x'}}{E'})}{\kmkd{i_x}{E}}{ES}{FS}\\
    \kunder{z}{(\kprim{-1}{veval}~V'~\kmkd{i'_{x'}}{E'})}{\kmkd{i_x}{E}}{ES}{FS} &\rightarrow& \textbf{let}~o = returnOk(z,i')~\textbf{in}\\
    &&dropRV(z,\kmkd{i_x}{E},ES)~\textbf{if}~o=\truev~\textbf{else}\\
    &&\kmkd{\katt{\emptyset}{\emptyset}}{(\kprim{-1}{veval}~V'~\kmkd{i'_{x'}}{E'})}\\
    \\
      \kpcombine{\kprim{0}{vau}}{(s'~(s\dots)~V)}{\kmkd{i_x}{E}}{ES}{FS} &\rightarrow& \kpval{\\&&\kmkd{genID()}{\kcomb{0}{s'}{\kmkd{i_f}{E}}{(s\dots)}{unval(V)}}\\&&}{\kmkd{i_{x}}{E}}{ES}{FS}\\
    \\
    \kpcombine{\kprim{0}{wrap}}{&&\\(\kmkd{i''}{\kcomb{n}{s'}{\kmkd{i'}{E'}}{(s\dots)}{V}})}{\kmkd{i}{E}}{ES}{FS} &\rightarrow& \kmkd{i''}{\kcomb{(S~n)}{s'}{\kmkd{i'}{E'}}{(s\dots)}{V}}\\
    \kpcombine{\kprim{0}{unwrap}}{&&\\(\kmkd{i''}{\kcomb{(S~n)}{s'}{\kmkd{i'}{E'}}{(s\dots)}{V}})}{\kmkd{i}{E}}{ES}{FS} &\rightarrow& \kmkd{i''}{\kcomb{n}{s'}{\kmkd{i'}{E'}}{(s\dots)}{V}}\\
    \\
    \kpcombine{\kprim{0}{if0}}{(V_c~V_t~V_e)}{\kmkd{i_x}{E}}{ES}{FS} &\rightarrow&
        \textbf{let}~F~=(\kprim{0}{if0}~V_c~V_t~V_e)\textbf{in}\\
      &&\textbf{let}~FS'~=\{F\} \cup FS~\textbf{in}\\
      &&\textbf{let}~V_c'~=\kpval{\kunval{V_c}}{\kmkd{i_x}{E}}{ES}{FS}~\textbf{in}\\
      &&\textbf{let}~V_t'~=\kunval{V_t}~\textbf{if}~F \in FS\\
      &&\qquad\quad\textbf{else}~\kpval{\kunval{V_t}}{\kmkd{i_x}{E}}{ES}{FS'}~\textbf{in}\\
      &&\textbf{let}~V_e'~=\kunval{V_e}~\textbf{if}~F \in FS\\
      &&\qquad\quad\textbf{else}~\kpval{\kunval{V_e}}{\kmkd{i_x}{E}}{ES}{FS'}~\textbf{in}\\
      &&\kunder{V_c'}{(\kprim{-1}{vif0}~V_c'~V_t'~V_e')}{\kmkd{i_x}{E}}{ES}{FS}\\
    \kunder{0}{(\kprim{-1}{vif0}~V'~V_t~V_e)}{\kmkd{i_x}{E}}{ES}{FS} &\rightarrow& \kpval{V_t'}{\kmkd{i_x}{E}}{ES}{FS}\\
    \kunder{(S~n)}{(\kprim{-1}{vif0}~V_c'~V_t'~V_e')}{\kmkd{i_x}{E}}{ES}{FS} &\rightarrow& \kpval{V_e'}{\kmkd{i_x}{E}}{ES}{FS}\\
    \kunder{z}{(\kprim{-1}{vif0}~V_c'~V_t'~V_e')}{\kmkd{i_x}{E}}{ES}{FS} &\rightarrow& \kmkd{\katt{\emptyset}{\emptyset}}{(\kprim{-1}{vif0}~z~V_t'~V_e')}\\
    &&\quad\text{(otherwise for if)}\\
    \\
    \kpcombine{\kprim{0}{int-to-symbol}}{(n)}{E}{ES}{FS} &\rightarrow& \kmkd{\emptyset}{'sn} ~\text{(a symbol made out of the number n)}\\
    \kpcombine{\kprim{0}{array}}{(V\dots)}{E}{ES}{FS} &\rightarrow& \kmkd{\kval}{(V\dots)}\\
  \end{array}
\]
  \caption{Semantics of Partial Eval Primitives}
  \label{fig:partialevalrimitives}
\end{figure}
\pagebreak

\section{Partial Evaluation Pseudocode}

\newcommand{\forcond}{$i=0$ \KwTo $n$}
\SetKwFunction{FRecurs}{PartialEval}%
\SetKwFunction{FRecursCall}{PartialEvalCall}%
\SetStartEndCondition{ }{}{}%
\SetKwProg{Fn}{def}{\string:}{}
\SetKwFunction{Range}{range}
\SetKw{KwTo}{in}\SetKwFor{For}{for}{\string:}{}%
\SetKwIF{If}{ElseIf}{Else}{if}{:}{elif}{else:}{}%
\SetKwFor{While}{while}{:}{fintq}%
\renewcommand{\forcond}{$i$ \KwTo\Range{$n$}}
\AlgoDontDisplayBlockMarkers\SetAlgoNoEnd\SetAlgoNoLine%

\begin{algorithm}[H]
\footnotesize
\SetAlgoLined
\Fn(\tcc*[h]{}){\FRecurs{form, env, env\_stack, evaluating\_forms\_set}}{
\KwData{Our inputs are the current form to partially evaluate,\\
the env to evaluate it in,\\
the full call stack of currently active envs,\\
and the set of evaluations currently taking place.}
\KwResult{The partially evaluated form}
\tcc{First we check to see if we will make any progress by partially evaluating this form}
$not\_yet\_evaled \leftarrow \truev~\in?~\kprog{x}$\;
  $newly\_real\_env\_ids \leftarrow \exists~i~\in \kprog{x}~\text{s.t.}~\kmkd{i_x}{E}~\in~env\_stack $\;
$newly\_unblocked \leftarrow evaluating\_forms\_set ~\cap~\kiprog{x}\neq~\kiprog{x}$\;

\uIf(){$not\_yet\_evaled \lor newly\_real\_env\_ids \lor newly\_unblocked$}{
    \uIf{form is an env value, $\kmkd{i_x}{E}$}{
        \tcc{grab the newer real version of this env, if it exists}
        \uIf{$\kmkd{i_r}{E'} \in env\_stack$}{
            return $\kmkd{i_r}{E'}$
        }
        \uElse{
            return $\kmkd{i_x}{E}$\;
        }
    }
    \uElseIf{form is an derived-combiner value, $\kmkd{i}{\kcomb{n}{s}{\kmkd{i'_x}{E}}{(s\dots)}{Tb}}$}{
        \tcc{recurse on the combiner's static env and body}
        \uIf{$x~=?~r$}{
          return $\kmkd{i}{\kcomb{n}{s}{\kmkd{i'_x}{E}}{(s\dots)}{Tb}}$\;
        }
        \uElse{
            $new\_env \leftarrow \kenv{(s \leftarrow \kmkd{i}{s})\dots}{s' \leftarrow \kmkd{i}{s'}}{env}$\;
            $new\_body \leftarrow \FRecurs{$Tb, new\_env, new\_env~\cup~env\_stack, evaluating\_forms\_set$}$\;
            return $\kcomb{n}{s}{env}{(s\dots)}{new\_body}$\;
        }
    }
    \Else{
        \tcc{this is a call - broken out into Algorithm 2}
        return \FRecursCall{form, env, env\_stack, evaluating\_forms\_set}\;
    }
}
\uElse {
    \tcc{partial evaluation won't make any progress, just return form unchanged}
    return form\;
}
}
  \caption{Partial Evaluation Algorithm Pseudocode (except calls)}\label{algo:partialeval}
\end{algorithm}

\pagebreak

\begin{algorithm}[H]
\footnotesize
\SetAlgoLined
\Fn(\tcc*[h]{}){\FRecursCall{form, environment, environment\_stack, evaluating\_forms\_set}}{
\KwData{Our inputs are the current form to partially evaluate,\\
the environment to evaluate it in,\\
the full call stack of currently active environments,\\
and the set of evaluations currently taking place.}
\KwResult{The partially evaluated form}

\tcc{first partially evaluate the combiner}
$\kmkd{x}{(T_1~T_2\dots)} \leftarrow form$\;
$c \leftarrow \FRecurs{$T_1, env, env\_stack, evaluating\_forms\_set$}$\;
\tcc{If the result is a suspended symbol lookup or a suspended call, return since we can't make progress}
\If{$c = \kmkd{x}{s}\lor c = \kmkd{\katt{x}{y}}{(T'\dots)}$}{
  return $\kmkd{\katt{\emptyset}{\emptyset}}{(c~T_2\dots)}$\;
}
\tcc{Otherwise, c is a combiner, either primitive or derived. Get its wrap level}

\uIf{$c = \kprim{n}{o}$}{
    $wrap\_level \leftarrow n$
}
\Else{
  $\kmkd{i}{\kcomb{n}{s'}{\kmkd{i'_{x'}}{E'}}{(s\dots)}{Tb}} \leftarrow c$\\
  $wrap\_level \leftarrow n$
}
    
\tcc{evaluate the the parameters until wrap\_level is 0 (or not at all if it is -1)}
$args \leftarrow (T_2\dots)$\\
\If{$wrap\_level > 0$}{
    $args \leftarrow ( \FRecurs{$T, env, env\_stack, evaluating\_forms\_set$}~\text{for}~T~\in args)$\;
    \If{any entry in args is not a value}{
        return $\kmkd{\katt{\emptyset}{\emptyset}}{(c~args\dots)}$\;
    }
    \Repeat{$wrap\_level = 0$}{
        $args \leftarrow ( unval(T)~\text{for}~T~\in args)$\;
        $args \leftarrow ( \FRecurs{$T, env, env\_stack, evaluating\_forms\_set$}~\text{for}~T~\in args)$\;
        \If{any entry in args is not a value}{
            $new\_c = replacewraplevel(c, wrap\_level)$\\
            return $\kmkd{\katt{\emptyset}{\emptyset}}{(new\_c~args\dots)}$\;
        }
    }
}
\uIf{$c = \kprim{n}{o}$}{
    return $drop\_redundant\_eval(o(env, env\_stack, args), env, env\_stack)$\;
}
\Else{
    \tcc{make our inner environment}
    $\kmkd{i}{\kcomb{n}{s'}{\kmkd{i'_{x'}}{E'}}{(s\dots)}{Tb}} \leftarrow c$\\
    $inner\_env \leftarrow \kmkd{i}{\kenv{(s \leftarrow V~\textbf{for}~(s,V) \in zip((s\dots), args)}{s' \leftarrow env}{\kmkd{i'_{x'}}{E'}}}$\;
    \tcc{Check if we're already evaluating this form, to prevent infinite recursion}
    $new\_c = replaceWrapLevel(c, wrap\_level)$\\
    \If{$(Tb,inner\_env) \in evaluating\_forms\_set$}{
        return $\kmkd{\katt{\emptyset}{(Tb,inner\_env)}}{(new\_c~args\dots)}$\;
    }
    $result \leftarrow \FRecurs{$Tb, inner\_env, \{inner\_env\}\cup env\_stack, \{(Tb,inner\_env)\} \cup evaluating\_forms\_set$}$\;
    \uIf{combiner\_return\_ok(result, env.id)} {
        return drop\_redundant\_eval(result, env, env\_stack)\;
    }
    \uElseIf{$s' \neq \emptyset$}{
        \tcc{If this combiner takes in the dynamic environment, track the current dynamic environment ID as needed for this call to progress}
        return $\kmkd{\katt{env.id}{\emptyset}}{(new\_c~args\dots)}$\;
    }
    \Else{
        return $\kmkd{\katt{\emptyset}{\emptyset}}{(new\_c~args\dots)}$\;
    }
}
}    
 \caption{Partial Evaluation Algorithm Pseudocode (for calls)}\label{algo:partialevalcall}
\end{algorithm}

\end{document}